\documentclass[12pt,preprint]{aastex}
\slugcomment{Accepted  for publication in 
 {\it The Astrophysical Journal}}  
\def\lax {\ifmmode{_<\atop^{\sim}}\else{${_<\atop^{\sim}}$}\fi}  
\def\gax {\ifmmode{_>\atop^{\sim}}\else{${_>\atop^{\sim}}$}\fi}  
\def\gtorder{\mathrel{\raise.3ex\hbox{$>$}\mkern-14mu
             \lower0.6ex\hbox{$\sim$}}}

\def\cm2{cm$^{-2}$}
\def\s1{s$^{-1}$}

\begin{document}

%\title{4U~1705-44:A Hybrid  {\it atoll} and {\it Z} Type X-ray Binary
% as a %(link between) 
%unique hybride type of atoll and {\it Z} source properties
%}
%\title{Stability of the photon indices in atoll-source 4U~1705-44 for spectral states 
%transitions
%}

\title{%{\it Beppo}SAX, {\it Suzaku} and {\it RXTE} 
X-Ray Spectra  of The High-Mass X-RAY Binary 4U~1700-37 using {\it Beppo}SAX, {\it Suzaku} and {\it RXTE} Observations
%. SPECTRAL HARDENING DURING the BANANA BRANCH
}

% Observational Evidence for Neutron Star in GX~340+0

%\title{The stability of spectral index of the ``hard component'' as a function of mass accretion rate in Z-source GX~340+0}
% Observational Evidence for Neutron Star in GX~340+0

%\title{On the Constancy of the Photon Index of  X-ray spectra of 4U~1728-34 through all spectral states} 
%during outburst transitions}

\author{  Elena Seifina\altaffilmark{1}, Lev Titarchuk\altaffilmark{2} \& Nikolai Shaposhnikov\altaffilmark{3} }
%\author{Elena Seifina\altaffilmark{1}, Lev Titarchuk\altaffilmark{2}}%  \& Filippo Frontera\altaffilmark{3} }
%\altaffiltext{1}{Moscow State University/Sternberg Astronomical Institute, Universitetsky 
%Prospect 13, Moscow, 119992, Russia; seif@sai.msu.ru}
\altaffiltext{1}{Moscow State University/Sternberg Astronomical Institute, Universitetsky 
Prospect 13, Moscow, 119992, Russia; seif@sai.msu.ru}
\altaffiltext{2}{Dipartimento di Fisica, Universit\`a di Ferrara, Via Saragat 1, I-44122 Ferrara, Italy, email:titarchuk@fe.infn.it}
%ICRANET, Piazza della Repubblica 10-12 65122 Pescara,  Italy; 
%  Goddard Space Flight Center, NASA,  code 663, Greenbelt  
%MD 20770, USA; email:lev@milkyway.gsfc.nasa.gov, USA}
%\altaffiltext{3}{NASA Goddard Space Flight Center, NASA, Astrophysics Science Division, Code 661, Greenbelt, MD 20771, USA; Chris.R.Shrader@nasa.gov}
%\altaffiltext{4}{Universities Space Research Association, 10211 Wincopin Cir, Suite 500, Columbia, MD 21044, USA}
\altaffiltext{3}{CRESST/University of Maryland, Department of Astronomy, College Park, MD 20742, USA} 
%Goddard Space Flight Center, NASA,  code 663, Greenbelt  
%MD 20771,  USA: email:nikolai.v.shaposhnikov@nasa.gov}
%\altaffiltext{3}{Dipartimento di Fisica, Universit\`a di Ferrara, Via Saragat 1, I-44122  Ferrara, Italy, email:frontera@fe.infn.it
%}

\begin{abstract}
We present  an X-ray spectral analysis of the high-mass 
%X-ray 
binary 4U~1700-37 during  its hard-soft state evolution. 
%between the  {hard}  and {soft} states. 
We use the {\it Beppo}SAX, {\it Suzaku} and {\it RXTE} ({\it Rossi} X-ray Timing Explorer) 
%({\it RXTE}), {\it Suzaku}  and {\it Beppo}SAX 
observations for this investigation. We argue that the X-ray broad-band  spectra during all  spectral states can be adequately reproduced by  a  
model, {\bf consisting} of  a low-temperature {Blackbody} component,  two Comptonized components both  due to the presence of a Compton cloud (CC) that up-scatters  seed photons {\bf of $T_{s1}\lax$1.4  keV, 
%coming from the compact object (presumably, neutron star),
%this is   the first component, {Comptb1}, 
and 
%seed photons of a color temperature 
$T_{s2}<$1 keV,} 
% coming from the disk (or extended accretion column),
%the transition layer (TL) %disk (CC ?)
%the second component {Comptb2} 
and an iron-line 
%({Gaussian}) 
component. 
%In addition, we detected a  cyclotron  absorption  line  at $\sim$ 36 -- 39  keV in {\it Beppo}SAX and {\it Suzaku} data. 
We find 
%Our spectral analysis 
using this model %indicates  
that the photon power-law index   is almost constant, $\Gamma_{1}\sim 2$ for all spectral states. {\bf  However}, 
$\Gamma_{2}$ shows a behavior {\bf depending on} the spectral state. Namely,    $\Gamma_{2}$ is quasi-constant at the level of $\Gamma_{2}\sim 2$ {\bf while the CC plasma temperature  $kT^{(2)}_e$ is  less than 40 keV; on the other hand,  
$\Gamma_{2}$ is    in the range of   $1.3<\Gamma_{2}<2$, when $kT^{(2)}_e$ is greater than 40 keV}. We explain this quasi-stability of $\Gamma$  during  most of hard-soft transitions of 4U~1700-37
in a framework of the model in which the resulting spectrum is described by two  Comptonized  components. We  find  that these Comptonized spectral components  
of the  HMXB 4U~1700-37 {\bf are similar to those    previously found in NS sources}.
% as the  {\it atoll} 4U~1728-34,  GX~3+1, 4U~1705-44 and 4U~1820-30 and  the {\it Z}-source GX~340+0 
% for which the compact object is a neutron star (NS).  
This  index dependence  versus both mass accretion rate and $kT_e$ {\bf  revealed in 4U~1700-37 %as in a number of other NSs, 
is a  universal  {observational} evidence for the presence of a NS in 4U 1700-37}.
\end{abstract}


\section{Introduction}

Our knowledge of diagnostics for a BH or a NS presence 
%as a compact object 
in  X-ray binaries has been challenged in the last years by growing observational 
evidence  that the  spectral index 
%of X-ray spectrum 
is uniquely sensitive to the 
type of the compact object (NS or BH ) during burst events (Titarchuk \& Shaposhnikov, 2005; Shaposhnikov \& Titarchuk 2009; Shrader et al. 2010;
Farinelli \& Titarchuk, 2011; Titarchuk et al. 2014, hereafter TSS14).
%references therein)
{\bf In other words, this observational signature can be used as a probe for the  nature of  a compact star}.  
 %This 
%diagnostical 
%method can be shortly formulated as an observational phenomenon during outburst.
The photon index $\Gamma$ of the energy spectrum tends to be  quasi-constant, 
%namely the photon index 
around $\Gamma\sim 2$ (or the spectral index $\alpha=\Gamma-1\sim1$) 
%during outburst progress in X-ray binary, 
for
% the compact object is, most probably,
 a neutron star. When  
%otherwise, if 
$\Gamma$  monotonically increases with an outburst flux and then saturates at the burst peak, 
%In that case 
this   compact object is identified as a { black hole (BH)}. The observational behavior of the index  in a NS  is combined  with the specific decrease of $\Gamma$, 
%if the $\Gamma$ additionally tends to drop 
from the level of $\Gamma=2$ to $\Gamma< 2$ 
at the high plasma temperatures of the transition layer (TL) when $kT_e>50$ keV
[see SST14 and \cite{STSS15}, hereafter STSS15].
% then the compact object is a NS. 
This  specific behavior  of $\Gamma$ versus $kT_e$ is never observed 
in  BHs and thus can be considered as a spectral signature of a NS.

The index  correlations for BHs were demonstrated   by  \cite{st09}, hereafter ST09.
%   and 
%those for NSs  were, for the first time,   shown by  
\cite{ST11}, 
%\cite{TSS14}
hereafter ST11,  discovered the photon index behavior versus mass accretion rate in NSs,   and then  it was confirmed by  TSS14.  
% (for the NS case). 
%Then it was observationally tested with BH diagnostics in application to both LMXBs and HMXBs. 
%However, 
Because the above NS diagnostic was only tested for low mass X-ray
binaries (LMXB), now we decided to check this index versus mass accretion rate
correlation also for high mass X-ray binaries (HMXB).
%The above NS diagnostic is 
% presented in literature 
%(e.g., ST11, ST12, STF13, TSF13, STS14, STSS15) 
%  only tested  for low mass X-ray binaries (LMXB).
%On the one hand, LMXBs are good laboratory for above NS test as they demonstarte 
%fair changes of %grossly changeable 
%spectral characteristics. 
Therefore, it is important to test %is 
this method, using this  index versus mass accretion correlation, 
 in application to  HMXBs which  are relatively young sources consisted  of 
a massive early-type stellar companion (OB/Be-type star)  
and a compact object (NS or BH).
%either a neutron star or a black hole. 
In some Be/X-ray binaries X-rays are produced when  matter accretes 
 to a compact object  passing through the dense equatorial disk of Be-star.
%Some of 
%In %the HMXBs %are 
%Be/X-ray binaries %, wherein 
%X-rays are produced only when a compact object 
%accretes matter by passing through the dense equatorial disk of Be-star. %However, 
Other HMXBs contain a more massive OB supergiant, which is  characterized by stellar 
winds with mass-loss rates of $10^{-7}$ to
$10^{-5}$ M$_{\odot}$ per year. A fraction of this outflow  may be
attracted by  a compact object. Usually, HMXBs with O-B stars produce 
{periodic}  X-ray outbursts in  a case  of high eccentric binary orbits during periastron passages. 
For the  circular orbit cases, %However, 
some of OB/X-ray binaries containing supergiants, 
%which overflow or close to  filling their Roche lobes (Petterson 1978), 
demonstrate large sporadic X-ray flares. % even for circular orbits.
%Accretion is probably  powered by 
Roche lobe overflow usually leads to an accretion disk and a wind.  
% from the disk and an accretion
%disk. 
Long-term observation revealed precession effects in V-optical light curve 
(see Khruzina \& Cherepashchyk, 1983) which can be interpreted as an indication of a disk component in  accretion flow.

%Even systems, in which 
%It is thought  that accretion
%dominates, are often close to filling their Roche lobes
%(Petterson 1978). 
%As the surface of the primary star
%approaches the critical potential surface, then it is presumably  a
%smooth transition between a stellar wind enhanced along
%the line between the stars and Roche lobe overflow (Friend
%\& Castor 1982).
% takes place through either Roche-lobe overflow or the strong stellar wind from the optical companion. 
%precisely this
%However, %in which case 
These binaries usually host black holes [e.g., Cyg~X-1 and a number of HMXBs in the Magellanic Clouds  and the Milky Way, see a review by \cite{lwvdk10}]. HMXB 4U~1700-37 is 
an interesting example of such OB/X-ray binary.
%The latter case is just realized in HMXB 4U~1700-37. 
However, the nature of the compact object in this system is still %opened to question 
an open question  (a NS or a BH?) in spite of the fact that 
observations of 4U~1700-37 have been obtained 
% in  wide  energy ranges 
for more than 40 years.

4U~1700-37 is a well studied accreting supergiant X-ray binary (SgXB) 
%which  was 
first detected
by the {\it Uhuru} satellite in 1970s (see Jones et al. 1973). A companion star is  O6.5 Iaf supergiant with a fairly
well measured mass and radius (Abubekerov 2004; Rubin et al. 1996; Heap \& Corcoran 1992). 4U~1700-37 displays 
regular eclipses every 3.41 days, which are 
%a periodicity of that has been 
firmly associated with the  orbital period 
(see Table~\ref{tab:table_param} for other binary parameters). 
This eclipse covers about 14\% of the orbit (i.e. 0.5 days), and residual emission is visible during this period probably 
%due to  
because of scattering of X-rays in a  wind from the supergiant star (Boroson et al. 2013). The compact object  associated with this system does  not exhibit
any pulsation, while Boroson et al. (2003) reported  an information  on quasi-periodic oscillations (QPOs). 
%Usually observations of a neutron star The lack of a clear detection of 
%pulsations in 4U~1700-37 is really surprising, since 
Most of NSs in X-ray binaries 
% HMXBs  
demonstrate  coherent pulsations [see e.g. \cite{fk98} that in 4U  1728-34]. Thus,  the origin of the compact 
object in 4U~1700-37 is not clear (see Gottwald et al. 1986; Clark et al. 2002). 
%POTOM
In comparisons  with other SgXBs, the  4U~1700-37 source displays  prominent X-ray long 
and  short variabilities  which are {\bf very similar to those seen, for example,  in Vela X-1 [see \cite{krey08}].} 
%that can be associated 
%in terms of  l variability. 
%on a daily (hourly) timescale (e.g., see Fig.~\ref{lc_1700}, $lower$ panel), which possibly caused by the presence of clumps in the wind. 
%It can be expected that that this system also undergoes short periods
%of disk-accretion, especially when the soft X-ray luminosity (which is proportional to mass accretion rate) is 
%significantly encreased (Fryxell \& Taam 1988, ApJ, 335, 86; Taam \& Fryxell 1988, ApJ, 327, L73; 
%Taam et al. 1988, ApJ, 331, L117, Romano et al. 2015, A\&A, 576, L4).  
Prominent   X-ray flares in 4U~1700-37 are observed by INTEGRAL for which flux increases  up to two orders of magnitude
(by a factor of $\sim$100), often reaching 1.5 Crab, or more (see Kuulkers et al. 2007). This can be   due to an increase of the rate of wind-fed accretion or due to magnetospheric storms.  That behavior of 4U~1700-37 is similar to that seen in previous observations by CGRO/BATSE, GRANAT/SIGMA
% as well as older experiments %;
%the lowest fluxes are reached during eclipse 
(Markert et al. 1979;  Pietsch et al. 1980; Laurent et al. 1992; Rubin et al. 1996; Kudryavtsev et al. 2001; Laycock et al. 2003; Orr et al. 2004).

Because of non-detection of pulsations and  the observed X-ray  hard tails, Brown et al. (1996) proposed that this source is a  BH candidate. However, Reynolds et al. 1999  found that the 2 -- 200 keV  {\it Beppo}SAX spectrum of 4U~1700-37 was  fitted by a high energy
cutoff power-law model which is  a typical spectrum of an accretion powered X-ray pulsars. %(Reynolds et al. 1999).
Moreover, using the {\it Beppo}SAX spectrum, Reynolds et al. (1999) also declared a
%possible 
presence of cyclotron absorption feature at $\sim$37 keV.
 
% In its turn, 
{\it Suzaku} observations  possibly revealed a pronounced cyclotron absorption feature at $\sim$39 keV and  spectral variability on ks time scales, due to both variable absorption around the average value  of $2\times 10^{22}$ cm$^{-2}$ and changes in the spectral slope (Jaisawal \& Naik, 2015). These results argue  against a possibility of a black hole 
%in a compact object 
in the 4U~1700-37 source. 

High resolution spectra of XMM-{\it Newton} and  {\it Chandra} observations were well fitted  by a two component absorption model which was previously utilized  for interpretation  of the  1991 April $GINGA$ observations (see Boroson et al. 2003 and van der Meer et al. 2005).
The {\it Chandra} observations during the intermittent flare state of 4U~1700-37 revealed 
 fluorescence emission lines from neutral atoms and the recombination lines from H and He like species. The line strengths   varied over the observations. Boroson et al. (2003) found  that  triplet structure in Mg  and Si  points to   the non-equilibrium  of photo-ionized plasma. 
%around X-ray source 
%(see Boroson et al. 2003). 
During eclipse  and low flux time intervals of the
binary  van der Meer et al. (2005) found many fluorescence emission lines  as well as  recombination lines  in the XMM-{\it Newton} spectrum  of 4U~1700-37.   An extended ionization region around the source was suggested because of presence of these recombination lines.
% from H and He-like  atoms in the eclipse phase. 
% Similar % to the
%case of 
%to accretion powered X-ray pulsars, 
Using the {\it Chandra} observations Boroson et al. (2003) also revealed a few of mHz QPOs
in the power density spectra   of 4U 1700-37.

We found a large range in the mass evaluations of the compact object
($M_X$) and secondary star ($M_V$) %, and their mass ratio ($q=M_x/M_v$)
in 4U~1700-37 (see Table 1). %The nature of the compact object was inferred using
%a mass estimate or its upper limit. However, 
Recent mass estimates of $M_X = 2.44  
\pm 0.27 M_{\odot}$ (Clark et al. 2002; Rude et al. 2010) are significantly above the usually 
accepted mass limit for neutron stars. It is worth noting that Abubekerov (2004) analyzed radial-velocity 
curves of hydrogen Balmer absorption lines 
for the  4U~1700-37 source 
%based on the Roche model 
using the 1973 -- 1976 IUE spectral data  (see Hammerschlang-Hensberge et al. 1978). They estimated  $M_X$
%mass of the compact object 
%in three  ways: 
(i) using the  current paradigm about the gravitational acceleration of the optical component 
($M_X\sim 2.2\pm 0.2$ M$_{\odot}$), (ii) using  a radius of the optical component 
($M_X\sim 1.7\pm 0.2$ M$_{\odot}$) and (iii) based on the mass-luminosity relation 
($M_X\sim 1.41\pm 0.08$ M$_{\odot}$). The upper limit of these estimates is in  agreement with the results 
of Dolan  (2011) in which $M_X\sim 2.44$ M$_{\odot}$ (see also similar estimates by Clark et al., 2002; Rude et al., 2010) and thus these estimates  
are significantly above the usually accepted mass limit for the neutron star.
The dynamical  measurement of this compact object  mass can be taken
%of more/less than the theoretical limit for a stable  BH/NS configuration 
%is currently considered 
as a   demonstration  that the object is a BH. However,
 such  a mass estimate is not a conclusive argument for the compact object nature.
%nature of an object. 
%as a NS, for example. 
A real proof should be followed from an observational corrobaration of the absence or presence 
of any solid surface for a particular object.
% which would otherwise be manifested by observations
%of a "response" in the form of a strong spectral component due
%to the energy release on the surface and/or coherent pulsations
%as in the case of neutron stars.

In this paper, we use an essentially new method for
diagnostic of the compact object origin 
%of the compact object 
in 4U~1700-37 applying  
an analysis of  available {\it RXTE}, {\it Beppo}SAX and  {\it Suzaku} 
observations. % of 4U~1700-37.
%In \S 2 we present 
The list of observations used for the data analysis is shown  in \S 2, while 
in \S 3 we describe  a reader  details of  our spectral analysis.  
We   discuss  how   X-ray  properties change  during the low$-$high spectral state  transition of 4U~1700-37 
%and present the results of the scaling analysis to estimate the mass of M101 ULX-1 
in \S 4. % and \S 5. 
%We consider an interpretation of observational results and show our arguments
%for NS presence in 4U~1700-37 in \S 5. %Section 6. 
We make our final  conclusions in \S 5.

\section{Details of observations and  reduction of the data \label{data}}

\subsection{Data Selection \label{data}}

\subsubsection{{\it Beppo}SAX}

%4U~1700-37 was observed by 
{\it Beppo}SAX in April of 1997 observed 4U~1700-37  during uneclipsed interval, $\varphi=0.45\div 0.59$ (see Table 1).  
We obtained the broad-band energy spectrum of the 
source using  the combined  data from  three {\it Beppo}SAX Narrow
Field Instruments (NFIs): the Low Energy Concentrator
Spectrometer [LECS; \citet{parmar97}], the Medium Energy Concentrator Spectrometer
[MECS; \citet{boel97}]  and  the Phoswich Detection System [PDS; \citet{fron97}] for  the 0.3 $-$ 4 keV, 1.8 $-$ 10 keV and   the 15$-$150 keV ranges respectively.
%NEW
%{
%the High Pressure Proportional Gas Scintillation Counter 
%[HPGSPC; Manzo et al. (1997)] for the 8 -- 50 keV range 
%}
%END NEW
% for the 15$-$150 keV range. 
%In this way, 
The source spectra  were extracted from the 
 event data %list 
using a circular region centered on the source of radius 30{\tt "}. 
While 
%the corresponding background spectra 
%from 
an annulus region within  30{\tt "}  and 180{\tt "} radii was used to determine  the background. 
We applied the SAXDAS data analysis package  for data reprocessing. 
We carried out the spectral analysis in the energy band
%for each of these {\it BeppoSAX} instruments,  
for which  the response  is well known. 
%The LECS data have been  renormalized based on MECS data. 
In a model fitting we treated a relative normalization of the NFIs as a free parameter while   
 the MECS normalization  was frozen at a value
 of 1. This method led to cross-normalization factors in the range, expected for each of these instruments  %Following  this  fitting procedure we checked 
%to ensure that 
% in order to check if 
%that these normalizations
% were in a standard range for each of the instrument
\footnote{http://heasarc.nasa.gov/docs/sax/abc/saxabc/saxabc.html}.
% Specifically, LECS/MECS re-normalization ratio is 0.92 and PDS/MECS
% re-normalization ratio is 0.97. 

The spectra were rebinned   according to  
 instrument energy resolution 
%of the instruments  
 to have
 % independent 
significant statistics in each bin.  Thus, the LECS spectra were rebinned using  an energy dependent  binning factor 
%which is not constant over energy 
(see \S 3.1.6 of Cookbook for the {\it Beppo}SAX NFI spectral analysis) and applying rebinning 
template files provided  by GRPPHA of  XSPEC\footnote{http://heasarc.gsfc.nasa.gov/FTP/sax/cal/responses/grouping}. 
The PDS spectra were rebinned with a  binning  factor 2, grouping two bins together (resulting bin width is 1 keV).  
For the $Beppo$SAX analyzed spectra we applied a systematic error of 1\%.  %thes
%END NEW
  % 1 
The {\it Beppo}SAX observations taken for our analysis are listed in  Table~\ref{tab_sax_suzaku}.

\subsubsection{\it Suzaku}

During the binary uneclipsed 
interval 
%of the binary 
spanning 0.29 -- 0.72 orbital phases 
 {\it Suzaku} in 2006, September 13$-$14 observed 4U~1700-37
 (see details in Rubin et al. 1996). 
The observations were implemented  in "XIS nominal" position using an effective exposure of $\sim$ 82.1 ks and $\sim$ 81.5 ks  for HXD and XIS, respectively. %During the observations  
XIS detectors worked  in the "burst" clock mode with "1/4 window" option providing  time resolution of 1 s.  
For the {\it Suzaku} observations we utilized publicly available data (version 2.0.6.13). 
% in the present work. 
 We used for our data analysis {\tt HEASOFT software package}
({\bf version 6.13}) and calibration database (CALDB) released on 2012 February 10  and 2011 September 13  for  XIS and HXD, respectively.
 We applied the unfiltered event files for each of the
operational XIS detectors (XIS0, 1 and 3) %; Koyama et al.2007) 
using the latest {\tt HEASOFT software package} (version 6.13)  and following  the {\it Suzaku} Data Reduction Guide\footnote{http://heasarc.gsfc.nasa.gov/docs/suzaku/analysis/}. %  using the latest HEASOFT software package (v6.13). % and 
 We obtained cleaned event files by re-running the {\it Suzaku} {\tt pipeline} implementing the
latest calibration database (CALDB) available
since 2013,  January 20  and also applying the associated screening criteria
files. 

Thus, we got the 4U~1700-37 spectra  from the 
filtered XIS event data %list 
taking a circular region, centered on the source, of radius 30{\tt "}, and the corresponding background spectra from an annulus region with 30{\tt "} and 180{\tt "}  radii.
  Using the {\it Beppo}SAX sample we considered  the background region to be in the vicinity  of the source extraction region. 
We  extracted spectra and lightcurves   from the cleaned event
files using XSELECT, and we generated  responses  for each
detector utilizing  the XISRESP script with a medium resolution.

Implementing  reprocessed HXD/PIN event file and
XSELECT package of FTOOLS  we created  HXD/PIN spectra and light curve of 4U~1700-37.
% were created   
  Applying
{tuned} non X-ray background (NXB\footnote{http://heasarc.nasa.gov/docs/suzaku/analysis/pinbgd.html}) event file we accumulated the HXD/PIN background light curves
and spectrum. 
%A correction for cosmic X-ray background (CXB\footnote{http://heasarc.nasa.gov/docs/suzaku/analysis/pin cxb.html}) 
We included  a correction for cosmic X-ray background (CXB\footnote{http://heasarc.nasa.gov/docs/suzaku/analysis/pin cxb.html}) 
%
%1 http://space.mit.edu/ASC/software/suzaku/aeattcor.sl
%2 http://space.mit.edu/ASC/software/suzaku/pile estimate.sl
%3 http://heasarc.nasa.gov/docs/suzaku/analysis/pinbgd.html
%4 http://heasarc.nasa.gov/docs/suzaku/analysis/pin cxb.html
%
in the PIN spectra as recommended %suggested 
by the instrument team.
%Epoch 2 response file (20080129) for HXD/PIN was used in the spectral analysis. 
%Data from all four XISs (XIS-0, XIS-1,
%XIS-2 and XIS-3) and HXD/PIN were used in the present study. 

In these {\it Suzaku} observations we have also found flare-type variability of a factor of 100 (XIS-0)  
on time scales from  minute to hours.
We assumed that {\bf the above X-ray variability} versus  orbital phase  within these observations %%, shown in the $bottom$ panel of Fig.~\ref{lc_1700},  
 was caused by a clumping wind structure or local $N_H$ variations. %(wind clump features) 
Note, that  {\bf this kind on inhomogeneity has already 
been observed in other SgXBs, like 4U 1538-52 and GX 301-2 [see \cite{rod15} and \cite{evan10}, respectively]}.

Thus, we did not relate these   X-ray variations  to 
mass accretion rate changes ($\dot M$).  
 A time averaged spectral analysis  was implemented using the HXD/PIN and XIS-0 spectra.
% from  and the .
We carried out  spectral fitting  applying XSPEC v12.7.1. 
 The energy ranges around of 1.75 and 2.23 keV are not  used for spectral fitting because of the known artificial structures in the XIS spectra around the Si and Au edges.
%Additionally, owing to 
%strong absorption, the spectrum of 4U~1700-37 has very limited statistics below 3 keV. 
 Therefore, for spectral fits we have chosen  the 0.3 -- 10 keV  range  for the XISs 
(excluding 1.75 and 2.23 keV points) and the 15 -- 70 keV  range
for the PIN spectrum. 
%We fitted the spectra
%simultaneously with all parameters tied, except the relative
%instrument normalizations which were kept free. 

\subsubsection{\it RXTE}

For our analysis, we  have also applied publicly available data of
the {\it RXTE} 
% \citep{bradt93}  
acquired  from 1996 September  to 2003 September. 
These data consist  of 62 observations related to   different spectral states of the source.
For data processing we  utilized standard tasks of the LHEASOFT/FTOOLS
5.3 software package.
Spectral analysis was implemented  using  PCA {Standard 2} mode data, collected 
in the 3$-$23~keV  range and  applying the most updated release of PCA response 
calibration (ftool pcarmf v11.7). We  also used the standard dead time correction 
 to the data and the data from HEXTE detectors 
%have also been used 
to make broad-band spectra.
A background collected  during  off-source observations was subtracted. For the spectral analysis  %using  HEXTE detectors 
we took only the HEXTE data  in the 19$-$200~keV  range. 
%in order to account for the uncertainties of the HEXTE response and 
%background determination. %, %The HEXTE data have been re-normalized based on the PCA.
 The GSFC public archive 
(http://heasarc.gsfc.nasa.gov) can provide these data. In  Table~\ref{tab:list_RXTE}
% \ref{tab_rxte} % 2 
we present a list of 
%the  groups
% of 
observations which covers the complete sample of the state evolution of the source. 
 %during different spectral  state events. 

  One can assess the source intensity  on month 
timescale using an   example of the  light curve (during  1996 September) of the All Sky Monitor (ASM, Levine et al. 1996) onboard of \textit{RXTE} shown in Figure \ref{lc_1999}.
 %The ASM data (during September 1996) indicate 
%that the source demonstrate 
%long term variability in  scale of month from 5 ASM counts/s (corresponding to the $hard$ state) to 35 ASM counts s$^{-1}$ 
%(the $soft$ state). Note that during the $hard$ state the source showed a type-I X-ray burst, while no bursts were 
%present in the $soft$ state of the source (D'Ai et al., 2010). In fact, we have  excluded the type-I bursts from the analysis.   Generally, the source presented  the hard-to-soft  transitions characterized by the different time scales of the X-ray flux variability. As seen from Fig.~1, we have chosen the time 
%intervals with %more 
%long (with high flux ``plateau'', $R1$ set) and with relatively shorter  time scale periods
%of variability ($R2$ and $R3$). 
As we have already mentioned above,  4U~1700-37 displays a prominent variability in the soft band  which can be associated with long-term variation (of order of ten days)
% Fig.~\ref{lc_1999}, upper panel) 
and  short-term variability 
on a daily (hourly) timescale (see e.g. see Kuulkers et al. 2007).
Extremely large sporadic outbursts %, well seen in $top$ panel of Fig.~\ref{lc_1700}, 
 indicate % It can be expected 
that this system  can be also affected by  %exposed by %undergoes 
short periods of disk-accretion, especially when  soft X-ray luminosity (proportional to 
$\dot M$)  
 increases significantly (Fryxell \& Taam, 1988; %Taam \& Fryxell, 1988; Taam et al., 1988; 
Romano et al. 2015). 
This type of  variability  of 4U~1700-37 is the main focus of the present investigation.  In order to do this,  we have   made an analysis of {\it RXTE} observations  of this source  spanning seven years  and 
divided them  to  five intervals indicated in Table~\ref{tab:list_RXTE}.
%\ref{tab_rxte}. %by  blue rectangles in Figure~\ref{variability_97-09} ($top$).
 We modeled the {\it RXTE}  spectra using XSPEC  fitting package and 
%{\bf The HEXTE spectra were renormalized to the PCA data by adding to the other free parameters a normalization factor?????} 
%%%Spectral analysis was done using an approach similar to that adopted in ST11 for 4U~1728-34 data. 
we applied a systematic uncertainty  of 0.5\%  to all  analyzed 
%{\it
%RXTE   
%}
%END NEW
spectra. 

\section{Results \label{results}}

As we have already pointed out above,  4U~1700-37 {\bf shows a complex X-ray behavior on a time scale of $\sim$ day probably caused 
by orbital modulation (Rubin et al., 1998)}. The source soft  photons  are also  subject to  a variable inner 
absorption related to  dipping 
%of it  
into  a stellar wind   during of orbital motion (Buff \& McCray, 1964; 
Haberl et al., 1989; Branduardi et al., 1978). Orbital variation of X-ray absorption can be additionally modified by gas stream (Haberl et al, 1989), wind interaction (Titarchuk et al., 2007; Bychkov \& Seifina, 1996a,b) 
and bow shock trailing the compact object (Livio et al., 1979; Blondin et al., 1990; 
Saraswat et al., 1996).  {\bf Stellar wind from supergiant stars are known to be inhomogeneous, because of clumping in the hot-star winds 
[see, for example, a model by \cite{hamann08}].}
%    can not be smooth and homogeneous, 
%&but rather filled  by dense clumps. 
The presence of these dense  clumps dramatically affects the mass loss regime  
and causes sporadic suborbital variability. %(see an exapmle with $Suzaku$ light curve in the bottom panel of Fig.~\ref{lc_1700}). 
Note, that aforementioned variabilities are not related to any global changes in $\dot M$. 
Indeed,  X-ray emission of 4U~1700-37 is modified by transitions between the spectral states (the low/hard and high/soft ones),  which are usually a result  of significant changes of $\dot M$ (see below).

Here, we focus  on a  type of variability and spectral %and timing 
properties of 4U~1700-37 related to the mass accretion rate changes. 
 %, which are %seen as the mild and slow variabilities.
Thus, we have only selected observations during the binary uneclipsed intervals
 of orbital period.  X-ray eclipse takes place around optical primary minima, {\bf in the  phase interval}  
$0.93<\varphi<1.07$  
%$|\varphi|\leq 0.07$ 
(see van Paradijs, Hammmerschlag-Hensberge \& Zuiderwijk 1978). {\bf To exclude 
the eclipse orbital modulation we have only utilized
observations occurred in interval $0.07\le \varphi\le 0.93$}.  {\bf This resulted 
% type of observations include
in 90 out-of-eclipse  intervals}. It is  worth noting that  27
observations occurred  during eclipses. 
% taken at different precessional and orbital phases were used for the spectral and timing analysis of orbital modulation effects.
%Orbital ephemerids were taken from Haberl et al. (1998).
 We applied the ephemeris of Rubin et al. (1996): orbital period $P_{orb} = 3.411581(27)$ d
and mid-eclipse T0 = JD~2~448~900.873(2) to calculate the orbital phase

\subsection{Hardness-intensity and Color-color diagrams of 4U~1700-37 \label{ccd}}

{\bf To evaluate the source spectral evolution in time we build hard (HC)
and soft colors (SC)} making 
%We  use   hard colors  and soft colors (SCs) to show shapes of  %demonstrate different 
%configurations of 
color-color diagrams (CCDs).
% in the spectral diagrams. 
Figure~\ref{CCD_HID_1700} shows 
%collected  
 CCDs and hardness-intensity diagrams (HIDs)  (see left panel and  {right panel}, respectively).
% of 4U~1705-44. 
CCD presents 
%ordinate and  abscissa 
 the count ratios:  hard color (20-40 keV/9-20 keV) versus soft color (4-9 keV/2-4 keV), 
while HID  demonstrates  hard color 
(20-40 keV/9-20 keV)  versus PCA %ASM 
count rate (2-40 keV). % measured in units of $10^{-9}$ erg/s/cm$^{-2}$.  
  ObsIds of the observational sets are indicated on the top of the right panel of Figure~\ref{CCD_HID_1700}. 
As we  show   our data sets related the low/hard to the high/soft states  cover different parts of the  CCD and HID.
%
%In Figure ~\ref{CCD_HID_1700} 
We also mark the data of the different sets as
% different colors.  
%The sets are indicated by different colors: 
{red} for  ObsId 10148-01-01-000, {blue}  for ObsId 10148-01-02-00, 
{bright~blue}  for  ObsId 10148-01-03-000, {green} for ObsId 30094-01-01-00, {crimson}  for  ObsId 30094-01-02-00, 
{ yellow}   for ObsId 30094-01-03-00 and {black} for ObsIs 30094-01-04-00, 30094-01-05-00.
%To avoid possible effects of secular shifts between different epochs %(the so-called parallel tracks) 
%we selected data sets observed in time intervals, which are not much far away distanced from each other.    %selected 

Data sets that we investigated are associated with different %($third$ and $fourth$) RXTE 
epoch. 
%It is clear  from this Figure 
Clearly  our  data do not indicate a  secular shift effect. 
The  CCD and HID form plain and smooth tracks.
% on the whole. 
%Furthermore, we test this effect using flux units instead of counts, 
%in which case the corresponding tracks ($hardness$-$flux$ diagrams, see Sect.~3.2.4 and 
%Fig. \ref{3HID}) also form plain and smooth track. 
Note, that the HMXB 4U~1700-37 is similar to LMXBs (Z-sources and atolls) in terms of their  CCDs and HIDs [see, for example \cite{STSS15}].
%while timing properties are quite different. The strong wind of donor star caauses the blurring 
%of X-ray spectrum.

%*****
As one can also see,  4U~1700-37 demonstrates a typical outburst behavior, marked
% when  the 
by CCD/HID  track evolution from %the quiescence to 
the LHS %and along the upper branch horizontal and over HIMS and SIMS 
to the HSS. %, see $harness-flux$ diagram (HFD) in Fig.~\ref{lc_1700}). 
%The source then returns to LHS along the lower horizontal branch and finally disappears into the quiescence.
The CCD and HID tracks are, however, an empirical description and thus one needs
 a physical interpretation of their behavior.
Therefore, we %select observations performed at different flux levels in the $hard$ and $soft$ states and 
intend to apply these data for a detailed X-ray analysis of the spectral  evolution of the source. %during its  state evolution. 
 % and proceed with spectral analysis. 

%333
\subsection{Details of Spectral analysis}
\subsubsection{Selection of the Spectral Model\label{model choice}}

%In view of ferocious 
{\bf Because of the debated orgin of the compact object in 4U~1700-47} (a NS or a BH),  we test 
the source spectrum using  different  models in application to $Suzaku$, $Beppo$SAX and $RXTE$ observations 
(see examples of the  source spectra in Figs.~\ref{spectrum_Suzaku_SAX}$-$\ref{rxte_spectra_one_two_comptb}).  We should point out that different Comptonization  models  are usually used for a BH (e.g., Stiele et al. 2013)  and 
%the second one  is typical 
for  a NS (e.g., Farinelli et al. 2003; Paizis et al. 2006).  Specifically, 
 we started  with a model of an absorbed bbody plus the Comptb. 
The line feature at 6 -- 7 keV range is 
approximated using the  Gaussian line profile.  This model [phabs*(bbody+\-comptb+gaussian)] is  successfully  applied for BH 
spectral modelling (see also, STS14).  In the case of 4U~1700-37, however  this model  poorly fits the data %both 
%a low e nergy part ($<$10 keV) % and high energy tail ($>$60 keV), 
as it is clearly seen from Figure~\ref{rxte_spectra_one_two_comptb}, where  we demonstrate  the 4U~1700-37  spectrum  
(ObsId=10148-01-02-00) observed by  {\it RXTE} on 1996, September 13  along  with the fit residuals $\Delta\chi$. 
On the {left} panel the model, which includes a single Comptb component, i.e. the phabs*(bbody+\-comptb+gaussian) model,  is applied. 
Clearly from this plot  the model  
gives unacceptable fit ($\chi^2_{red} = 12.3$ for 86 dof). Significant positive residuals at high energies ($E>60$ keV) indicate  the existence of an additional emission component. As a result, {\bf we also add  
another   Comptonized  component} (Comptb2)  with a different seed photon temperature ($kT^{(2)}_s$) which results 
% Note, that  Fig.~\ref{rxte_spectra_one_two_comptb} ({left} panel) demonstrates the 
%unsatisfactory fit  
%the source spectrum 
%in the  30 -- 40 keV range. Therefore, we also use 
%a  cyclotron absorption profile, (cyclabs model), see Makishima et al. (1990) to account for these residuals. 
%The inclusion of %the cyclabs model along with 
%the second comptb component  
 in a significant  improvement  of the fit (see below).
%$\chi_{red}^2$.

In Figure \ref{rxte_spectra_one_two_comptb} (right panel) we  demonstrate the best-fit spectrum,  
%the best-fit spectrum 
and $\Delta\chi$ for this fit where the model is: 
%phabs*cyclabs*(bbody+\-comptb1+comptb2+gauss), 
phabs*(bbody+\-comptb1+comptb2+gauss).
This model  includes two Comptb components %and a cyclotron line component 
and $\chi^2_{red}$ = 1.12 for 81 dof.
We show the data by crosses and the  spectral  model %  {\it wabs*(blackbody+Comptb1+Comptb2+Gaussian)} 
is presented  by light-blue line. {\bf The components of the model   are color-coded} in red, green, dark-blue 
and crimson lines for  {Comptb1}, {Comptb2}, {Blackbody}  and {Gaussian} components, respectively. 

We have also  applied various models to  the available {\it Beppo}SAX and {\it Suzaku} data  using  a better spectral coverage and detector resolution 
 for the 0.3$-$200 keV range  %broad band spectra with 
%for different models 
%in the similar manner 
(see Table~\ref{tab:fit_table_SAX_Suzaku}). 
In Figure~\ref{spectrum_Suzaku_SAX} we show %present  
%two representative
 the {{\it Beppo}SAX ({left}, ObsId=20339001) and {\it Suzaku} (right}, ObsId=401058010) spectra of 4U~1700-37 
%along with
%the best-fit 
fitted using the two-Comptb model. % model $phabs*cyc*(bbody + comptb + comptb + gaussian)$. 
The data  are shown by crosses and the best-fit spectral  model,  phabs*(bbody + comptb + \-comptb + gaussian)  
%phabs*cyclabs*(bbody + comptb + \-comptb + gaussian) 
% {\it phabs*(bbody+Comptb1+Comptb2+Gaussian)} 
by light-blue line. The model components  are also  presented by  red, green, dark-blue and crimson lines for  
{ Comptb1}, {Comptb2}, { Blackbody}  and {Gaussian} line, respectively. 
{$\Delta \chi$ versus photon energy in keV is shown in the bottom panels. 
The best-fit parameters of the model  for the {\it BeppoSAX} data %HB 
 (see {left} panel) are 
$\Gamma_1$=2.00$\pm$0.03, $kT^{(1)}_e$=15.3$\pm$0.4 keV, $\Gamma_2$=1.41$\pm$0.02, $kT^{(2)}_e$=96$\pm$8 keV, 
$kT_{BB}$=0.56$\pm 0.07$ keV  and $E_{line}$=6.48$\pm$0.07 keV ($\chi_{red}^2$=0.96 for 335 dof). 
%The best-fit energy, $E_{cyc}$, and width, $\sigma_{cyc}$, of the cyclotron absorption feature are 
%$36.7\pm 0.9$ keV and $11.95\pm 0.04$ keV, respectively. 
Notably, %using  the same observation 
Reynolds et al. (1999)  fitted the same data by an absorbed power-law model, combined
with a high energy cutoff,  low energy thermal bremsstrahlung components, %also 
 an iron line and cyclotron absorption component. 
While this  model is often used for accreting pulsar spectra fitting,  it provides  almost the same power-law index of 
2.07$\pm$0.02 as that in %the case of 
 our model for the fist Comptonized component, Comptb1. Using the second Comptonized component of our  model, Comptb2, 
we found  that the hard tail in 4U~1700-37 spectrum is similar to one observed in 4U 1705-44 (see STSS15). 

%It is interesting that for the time-averaged $Suzaku$ spectrum 
%we also detected some dip like residuals in the PIN data. Therefore, we also used cyclotron absorption profiles 
%{cyclabs} to account for these residuals. 
%The best-fit cyclotron  line energy, $E_{cyc}$, 
% and width, $\sigma_{cyc}$, of the cyclotron absorption feature are 
%$39.6\pm 0.2$ keV and $8.9\pm 0.7$ keV, respectively. 

%34343

 { \bf As we have already mentioned,
 Jaisawal \& Naik (2015) applied  another model to fit  the $Suzaku$ spectra.
In particular, they 
used  a combination of partial covering NPEX, %partial covering 
a high energy cutoff power-law model,   
three  Gaussian-line  and cyclotron absorption components. As a result, they claimed a detection  of  %they  detected  
a cyclotron line at $\sim 39$ keV.  Using {\it Beppo}SAX data Reynolds et al. (1999) also suggested an existence of the cyclotron absorption feature at $\sim$37 keV
in the spectrum. Keeping in mind  these results, we have also tested the presence of this %absorption cyclotron 
line centered at 
$\sim$ 36 -- 39 keV using these available $Beppo$SAX and $Suzaku$ spectral data and our spectral  model}. While this model is statistically valid but its performace is worse than that using our model (see Table 4). 

{\bf  More specifically, we fitted these spectral data using
$phabs*cyclabs*(bbody+comptb1+comptb2+gauss)$ 
model. For the  BeppoSAX spectrum (ObsId=20339001) we found $\chi_{red}^2= 1.17$(335
dof), while for the $Suzaku$ one  (ObsId=401058010) we obtained $\chi_{red}^2= 1.21$(416 dof). 

%We should admit that these spectral fits are also acceptable but they are worse than those using  our  model 

In fact, one cannot see any  structure near $\sim$40 keV   in the residuals of our spectra    
(see Figure \ref{spectrum_Suzaku_SAX}). Thus, we are not able to   support the presence of this line  in the  $Beppo$SAX and Suzaku spectra using our spectral model}.

\subsubsection{Modelling of 4U~1700-37 spectra with two-Comptonization components}\label{spectral analysis}

The best-fit models  for   {\it RXTE},  {\it Beppo}SAX and {\it Suzaku} spectra 
were obtained %a so called 
using the {two-COMPTB} model phabs*(Bbody+Comptb1+Comptb2+Gauss).
% which we use in further analysis. Advantageously, 
In fact, this model provides a  physical picture, in contrast to previously 
models applied to X-ray spectra of 4U~1700-37 (see e.g., Reynolds et al. 1999; van der Meer et al. 2005; Jaisawal \& Naik, 2015). %(see Table~\ref{tab:BeppoSAX_fit_table}).
We want to emphasize that  the two-COMPTB model is usually applied to NS spectra, 
because in NSs,   the two Comptonized components are created in {\bf the CC} for which the seed photons come from the NS surface and the accretion disk.  %the one from TL area and  the second from NS surface with different seed soft temperatures. 
On the other hand, in  a BH  
%because of the absence of solid surface of the central sources (BH)
the seed soft  photons are only produced  in the accretion disk which is located next  to the relatively hot  CC.
% (see Fig. \ref{geometry}).

Our model for 4U~1700-37  describes a scenario 
%related to our model 
in which an accretion disk 
%is followed by the Compton cloud and then it  
is connected to the compact object (NS) 
%{\it Neutron star} 
through  the Compton cloud (CC) (see a possible geometry figure in STSS15).  
%)].
%Our model for 4U 1700-37 describes a scenario in which the system 4U~1700-37 is
%composed of an O6f star 
%having a strong wind, and an accreting neutron star. 
Note,  X-ray flux is highly
variable (by factor $\sim$ 100), 
showing evidence for a clumpy stellar wind (Jaisawal \& Naik 2015), which probably
results in a varying
accretion rate. 
%Because we detected the cyclotron feature at the energies of 40 keV, which
%is a signature of the presence of  a strong magnetic field we do not exclude a possibility that 
%with $\sim 10^{12}$. Since NS in high-mass binary 4U~1700-37 has strong field, the
%field is able 
%to capture and channel matter. Therefore, 
%accreting matter flows along field lines to 
%connect to 
%the magnetic  polar regions (see Fig. \ref{geometry}). 
%Therefore, most of the accreting matter falls on a region which is a
%fraction of 
%the whole surface area of the star. 
%As a result, the accretion energy can be released 
%in the accretion column or in the Compton cloud (CC).

 % at the inner edge thereof, which is innermost to NS surface.  %(see Fig.~2).
%In Figure~\ref{geometry}  we illustrate  our spectral model.
%We assume that accretion into a neutron star takes place when the material passes through 
%the two main regions:  a geometrically thin accretion disk [for example, the standard Shakura-Sunyaev 
%disk, see \citet{ss73}]
%and the TL,
% transition layer (TL), 
%where the soft NS surface  and  disk photons   are  upscattered off hot electrons of the TL. 
In our picture, the emergent thermal Comptonization spectrum is  formed in  the hot Compton cloud (CC), 
where soft photons of temperature $kT_{s1}$ from the neutron star  and the accretion disk of temperature $kT_{s2}$ are up-scattered  off the  hot  electrons 
%{\bf 
 giving rise to two 
%the  Comptonized 
components, {Comptb1} and {Comptb2}, respectively.
%} 
Some fraction of these seed soft photons can be also observed  directly from
 by the Earth,  
%{\bf 
that explains  why we add 
a soft blackbody  of the temperature $T_{BB}$ with   normalization $N_{BB}$.
%} 
%Red and blue photon trajectories, shown in Fig.~\ref{geometry},  correspond to soft 
%(seed) and hard (upscattered) photons, respectively. 
 Note,  the {Comptb} model describes the resulting spectrum as a convolution 
of a seed blackbody of
normalization $N_{Com}$ and temperature $kT_s$
 with the Comptonization  Green function.

%It is important to emphasize that 
The spectral index  of the emergent spectra of the Comptonization component 
% {\it Comptb1} and {\it Comptb2}  
is  determined by the energy release in the CC. As 
TSS14 demonstrate that  if the  gravitational energy release occurs throughout  the  cloud  around a NS then the spectral index should be around 1 (or the photon index $\Gamma$ is  around 2). However, if the  release of the energy  takes place only in  outer portion of the CC 
then the spectral index $\alpha$ is less than 1 (or  $\Gamma$ is less than 2). 
The latter case is realized in the sources for which  the local emergent luminosity exceeds  the Eddington limit and where the plasma  temperature of  the CC outer part is higher than 30 keV. Note, that in this particular case %of the  high plasma temperature of the CC 
the critical luminosity  can be a factor of 1.2-1.4  higher than the Eddington one depending 
on  a value of $kT_e$ (see TSS14 and Figure 10 there). This effect of the index decrease  has been  established in Sco X-1 and  4U 1705-44 (see TSS14 and STSS15,  respectively), which are   characterized by high luminosites. Our goal is to investigate this effect in 4U 1700-37.

For  the  XSPEC {Bbody}  model,
normalization $N_{Com}$ is 
\begin{equation}
N_{Com}=\biggl(\frac{L}{10^{39}\mathrm{erg/s}}\biggr)\biggl(\frac{10\,\mathrm{kpc}}{D}\biggr)^2.
\label{comptb_norm}
\end{equation}  
where $L$ is source luminosity of a source and $D$ is  distance to the source.
%In result, 
The free parameters  of the applied model, phabs*(Bbody+Comptb1+\-Comptb2+Gauss), are:
%To summarize the resulting spectral model parameters are 
the equivalent hydrogen absorption column density $N_H$; the  spectral indices $\alpha_1$, $\alpha_2$;
% (photon index $\Gamma=\alpha+1$); 
the {seed} photon temperatures %color temperatures of the $Blackbody$-like photon spectra 
$T_{s1}$, $T_{s2}$; $\log(A_1)$ and  $\log(A_2)$  are linked to the Comptonized
fractions $f_1$, $f_2$  ($f=A/(1+A)$);
 %is   the relative weight of the Comptonization component; 
the plasma temperatures $T^{(1)}_e$ and  $T^{(2)}_e$;  normalizations of $N_{Com1}$ and $N_{Com2}$ of the Comptb1 and Comptb2, respectively. 
%  The $COMPTB$ spectral component has the following parameters:
%temperature of the seed photons $T_s$,
%energy index of the Comptonization spectrum $\alpha$ ($=\Gamma-1$), 
%electron temperature $T_e$,   Comptonization  fraction $f$ [$f=A/(1+A)$, which is the relative weight of the Comptonization component] 
%and the normalization of the seed photon spectrum $N_{COMPTB}$. 
%We also include a soft blackbody component of the temperature 
 % $T_{BB}$ with  the normalization $N_{BB}$. 
 %was included. 
%
%Furthermore,  to the above model.  
In the model we also include a XSPEC {Gaussian},   whose   parameters are % a centroid line energy 
$E_{line}$, the line  width  $\sigma_{line}$  and line  normalization $N_{line}$,  to fit the data in the 6$-$8 keV  range.  
%To account for the residuals in 30 -- 40 keV range we also use 
% cyclotron absorption profile {cyclabs} described  by the model parameters as  a cyclotron energy $E_{cyc}$,  width of the fundamental line $\sigma_{cyc}$ and depth of the fundamental line $D$.

%(see Fig. \ref{BeppoSAX_spectra}). 
%We also include  the interstellar absorption with a column density $N_H$ in the model.

% LLLLLLLLL
Note,  we  fixed the following  $Comptb$ model parameters: 
$\gamma=3$
% related to the index of the low energy part of the spectrum, namely   
%$\alpha_{1/2}=\gamma_{1/2}-1=2$, 
%(low energy index of the seed photon spectrum) 
and $\delta=0$  [see meaning of these parameters in \cite{TSS14}).
%because we neglect the efficiency  of the  bulk inflow effect versus  the  thermal Comptonization   
%for  accretion into  a NS.
%  Sco~X-1.  
%The bulk inflow  Comptonization should take place very close to a NS surface, however, if the 
%radiation pressure  is high then the bulk motion is suppressed. On the other hand. if mass accretion  
%is quite low then the effect of the bulk motion 
%the bulk effect motion effect 
%is negligible too. Generally, the bulk effect in NSs is a  rare event. 
%We  also use a fixed value of hydrogen column $N_H=3\times 10^{21}$ cm$^{-2}$, which was found by Christian \& Swank (1997). %~\cite{iaria06}.  
%In addition, the parameter $\log(A_2)$ was fixed  at a value of 2, on the assumption %because 
%that NS  surface is completely covered  by TL (see Fig.~\ref{geometry} for 
%the Comptonization model geometry). 
% which provide high Comptonization plasma of NS surface. 
%When parameter $\log(A_2)\gg1$, this parameter is fixed at 2.0. 
 The value of  $\log(A)$ is frozen at 2 when $\log(A)\gg1$  because the model fit becomes insensitive  to the parameter 
%in any case of $\log(A)\gg1$ 
 $f = A/(1 + A)$.
% approaches unity 
%and equals approximately
%to 1 
%and further variations of $A\gg1$ do not improve fit quality any more.
% see Table 4). 
%We  use a value of hydrogen column $N_H=5.7\times 10^{22}$ cm$^{-2}$, which was found by $Beppo$SAX and $Suzaku$ data ( %~\cite{Barret_Olive02}.  
%In addition, the parameter $\log(A_2)$ was fixed  at a value of 2, when the best-fit values of $\log(A_2)\gg1$ because
%in any case of $\log(A_2)\gg1$ a Comptonization fraction $f = A/(1 + A)$ approaches unity 
%and equals approximately
%to 1 
%and further variations of $A\gg1$ do not improve fit quality any more. 
%As was shown in Sect.\ref{model choice} than %the $double$-$Comptb$ model 
%the X-ray broad-band energy spectra for {\it Beppo}SAX and {\it Suzaku} observations %during all %these 
%spectral states 
%can be adequately reproduced by  a %{\bf physical} 
%model, composed % sum % composition 
%of  a low-temperature $Blackbody$ component,  two Comptonized %({\it COMPTB}) 
%components (both  due to the presence of a TL that up-scatters both seed photons of $T_{s1}\le$1.1 %lax$
% keV coming from the compact object (neutron star or black hole)  (first component {\it Comptb1}), 
%and seed photons of temperature $T_{s2}\sim$1.3 keV coming from the transition layer (TL) %disk (CC ?)
%(second component {\it Comptb2}) 
%and the iron-line ({\it Gaussian}) component.
%In addition, we detected a  cyclotron  absorption  line  at $\sim$ 36 -- 39  keV in this spectra. 
We list all best-fit parameters in Table~\ref{tab:fit_table_SAX_Suzaku}.
% and Figure~\ref{spectrum_Suzaku_SAX}. 

The most  important issue is how  the photon index behaves depending on the different model parameters. These dependences are crucial for understanding   the compact object origin  in  4U~1700-37.  In this respect, for the $Suzaku$ and {\it Beppo}SAX  data 
%observations 
(see Table \ref{tab:fit_table_SAX_Suzaku}) we find that the spectral indices $\alpha_1$ and $\alpha_2$ are around  1.00$\pm$0.02 and 0.34$\pm$0.08, respectively 
(or 
%the corresponding photon indicies 
$\Gamma_1=\alpha+1$ and $\Gamma_2 = \alpha_2+1$ are  %about 2
 2.02$\pm$0.02 and 1.34$\pm$0.08, respectively).
%for $double~COMPTB$ model). 
We also reveal that $kT_{s1}$ and $kT_{s2}$ change  in the intervals of 1.3$-$1.4 keV and 
%about 0.8 keV
0.8$-$0.84 keV, correspondingly,   for all available %{\it Beppo}SAX 
data sets. 
Thus, using $Suzaku$ and $Beppo$SAX  data we reveal 
%that there are 
two  main blackbody sources: the first one is associated with the NS surface as 
the second one is presumably associated with the accretion disk region
% and the third %another one is related to the CC region,
for which 
%temperatures of soft photons are about %0.7 keV 
$kT_{s1}\sim1.3$ keV and $kT_{s2}\sim 0.8$ keV, respectively (see Table 4).

%\subsubsection{{\it RXTE} data analysis}\label{RXTE data analysis}

%3535

{\it RXTE} detectors can provide data   only above 3 keV, while %  but  
using the broad energy band of  {\it Suzaku} and {\it Beppo}SAX we can find    
 the low energy blackbody  parameters. 
%of {blackbody} components  at low energies.  
Thus, for the {\it RXTE} spectra  we should  fix   the blackbody temperature at a value of 
$kT_{BB}=$0.6 keV obtained using our   analysis of the $Suzaku$ and $Beppo$SAX data.
 In Tables~\ref{tab:fit_table_RXTE_1} and \ref{tab:fit_table_RXTE_2} we report the  best-fit spectral parameters of the two-Comptb model as a result of  analysis of  the {\it RXTE}  observations.   .  
 In Figure~\ref{6_spectra_rxte} we show six spectral evolution diagrams 
%for the hard  and soft states
 (see upper and lower panels). The RXTE data  (denoted by crosses) correspond to  observations, 30094-01-01-10 (green); 3009401-12-00 (orange); 30095-02-02-20 (violet); 30094-01-33-00 (red),  30094-01-31-00 (blue) and 30095-01-01-00 (black).   
%The spectral model components are displayed  by dashed $red$, $green$, 
%$blue$ and $purple$ lines for $Comptb1$, $Comptb2$, {\it Blackbody} and {\it Gaussian}, respectively.
Moreover, from Figure~\ref{6_spectra_rxte} %this  Figure %\ref{sp_compar_xte} 
we can  establish how the  spectral shape changes  in the energies  greater than 30 keV reflecting  an effect  of a sum of the  two Comptonized components for different   states.
 The hard tail at 50$-$200 keV  grows   with X-ray luminosity, except for the {\it intermediate soft} state [see 30094-01-33-00 spectrum (red)].
In our data  we have found many spectra with the strong
% a lot of in which 
 {high energy tail}  which extended to 200 keV  (see  Figure~\ref{6_spectra_rxte}).

As one can see from Tables 4-6  
% We find that
 normalization %electron temperature $kT^({1})_e$ 
of the  Comptb1 component changes in the range  of 0.01 to 22.6 in units of $L_{37}/{D^2}_{10}$ (where $L_{37}$ is the seed blackbody  luminosity 
 in units $10^{37}$ erg sec$^{-1}$ and $D_{10}$ is distance in units of 10 kpc),  
while the photon   index $\Gamma_1$  is almost constant ($\Gamma_1=1.99\pm 0.06$) %, $\Gamma_2 = ???$,)
 for all set of observations (see Fig.~\ref{two_state_spectra}, left diagram). 
%It is interesting that
 However, we  established  a  
two-phase sample for $\Gamma_2$: the phase of the quasi-constancy of  $\Gamma_2 = 2.01\pm 0.07$ when 
$kT^{(2)}_e$ changes in the 3$-$40 keV interval; and the low photon index phase of   $\Gamma_2<2$ for 
%the high electron temperature 
$kT^{(2)}_e>40$ keV (see Figure~\ref{two_state_spectra}, right panel).   Note that the  Gaussian line width $\sigma_{line}$   does not  change much  and numerous fits  indicate that $\sigma_{line}$ varies  in the  0.5 -- 0.8 keV range.
%Furthermore, a detailed analysis XMM-$Newton$ spectra of 4U~1705-44  recently carried out   
%by \cite{D'Ai10}, \cite{diSalvo09}, \cite{Fiocchi07} and \cite{egron13} has revealed that the iron line is quite broad during all spectral states.
% the line profile does not show any strong correlated variation for HB.  
Thus, $\sigma_{line}$ has been fixed at 0.7 keV 
%for all spectra 
for all $RXTE$ spectral fits. The plasma temperatures $kT^{(1)}_{e}$ and $kT^{(2)}_{e}$ 
%of the Comptb1 and Comptb2 components 
change in broad ranges of  2$-$22 keV and 19$-$100 keV, respectively (see Tables 4-6 and Figure~\ref{two_state_spectra}).
%~\ref{index_temperature_12}).%from  3 keV to 21 keV, while 
%the electron temperature $kT^{(2)}_{e}$ of $Comptb2$ component changes from  3 keV to 100 keV. 
%To follow up particular 
%evolution of spectral parameters along transition between $hard$ ans $soft$ states we it is convenient to use 
%{\it hardness-flux diagram} for 4U~1700-37.

%{spectrum_Suzaku_SAX}
%\subsection{The spectral features/evidance of BH source in 4U~1700-37}
%\subsubsection{The spectral model composition indicate on BH nature of X-ray source}

%222
\subsection{Overall picture of X-ray properties}

%To understand what are possible reasons which can cause the above mentioned hardening of 4U~1700-37 spectrum, %at UB, 
In Figure \ref{two_state_spectra} we also plot the photon indices $\Gamma_{1}$ (blue) and $\Gamma_{2}$ (red)  versus
%the normalizations 
the seed photon temperatures $T_{s1}$, $T_{s2}$ which 
% of the Comptb1 and Comptb2 components, respectively,  %measured in keV %$L_{39}/D^2_{10}$ units 
% in the frame of  our spectral model $wabs*(blackbody+Comptb1+Comptb2+Gaussian)$ during {\it RXTE} observations 
%outburst transitions 
%(see  Figure \ref{two_state_spectra}). 
%Blue and red points correspond to Comptb1 and Comptb2 components which are 
related  to the up-scattering  of  the NS and  accretion disk soft  photons. 
%of the NS surface and of the disk
%respectively. 
%by the plasma of the TL.
%(TL or CC).  
%at  the NS boundary layer and in CC, respectively. 
As one can see from this Figure that  $\Gamma_{2}<2$  when the $kT_{s2}$ decreases from 1.1 to 0.8 keV. {\bf This  is the strong demonstration} that $\Gamma_{2}$ drops when the Compton cloud area  enlarges. 
%On the other hand in the right panel of Figure~\ref{two_state_spectra} we show the photon indices $\Gamma_1$ and $\Gamma_2$ plotted %vs 
%versus the best-fit electron temperatures of the $Comptb1$ and $Comptb2$ components
%(see also Table 5).  $Blue$ and $red$ points correspond to $Comptb1$ and $Comptb2$ components. 
%which are related to 
%thermal upscattering of soft  photons by plasma electrons in  NS boundary layer and CC, respectively. 
% It is clear 
%seen that 
As we have already pointed out the { diminishing index phase} 
%of the Comptonization  component 
corresponds to the high plasma temperature phase  ($kT^{{(2)}}_e>40$ keV) while 
%the {\it index constancy} occurs 
for %the electron temperatures 
$kT^{{(2)}}_e$  in the 2$-$30 keV range   the index remains constant.
% associated with  another, relatively lower, temperature range (3 keV $<kT^{{(2)}}_e<85$ keV) of the same component.
%\label{index_temperature_s_12}

\bf This two component Comptonization model gives very good fits for 
all investigated data sets. From  Figures~\ref{spectrum_Suzaku_SAX} -- \ref{6_spectra_rxte}  one can see that the source spectra  are well fitted  by our two$-$Comptb model: 
$\chi^2_{red}\sim 1$
%where $N_{dof}$ is a number of degree of freedom, 
 for most of the cases. For 2\%  of the observations corresponding to higher S/N  
  $\chi^2_{red}$ reaches 1.4}. 
%However, it never exceeds a rejection 
%limit of 1.5 (for 90\% confidence level). 

%%

From top to bottom in Figure~\ref{lc_1999}  we show  the  {\it RXTE}/ASM count rate  evolution {\bf during the R2 1999 observations}, 
%{\it From Top to Bottom:}
%Evolutions of  count rate [2-9 keV] in counts s$^{-1}$ with 16~s time resolution, 
the   (10-50 keV)/(3-10 keV) ratio ($pink$) and the (50-200 keV)/(3-10 keV) ratio ($green$), 
%the  model flux in the 3-10 keV and 10-50 keV energy ranges ($black$ and $green$ points, respectively),  
 $kT^{(1)}_e$ ($red$) and $kT^{(2)}_e$ ($blue$) 
%of the $Comptb1$ and $Comptb2$ components, respectively, 
in keV,  
normalizations of Comptb1, Comptb2 and Bbody components (red, blue and black, respectively),    
and  the indices $\alpha_1$ and $\alpha_2$ 
%($\alpha_{1,2}=\Gamma_{1,2}-1$) 
(red and blue, correspondingly).
% for  1999 evolution %
%events ($R2$ set). 
%Particularly, 
Notably,  the Bbody component is very weak, sometimes it is detected at the limit of ``visibility". However,  for a number outburst states we detected the strengthening of this  component.
Blue vertical strips denote  the light curve phases when the  Bbody component increases.
%(based on timing analysis) 
%to the {\it reduced spectral index $\alpha_2$}, %flaring branch}, 
%to the {increased Bbody component}, %flaring branch}, 
%are marked with . 
%\subsubsection{Quasi-constancy of the Photon Index at the sub-Eddington
%Regime and a Reduction of the Photon Index $\Gamma_2$ at the High Temperature State}
The photon index $\Gamma_2$
% related to the TL 
drops when  $kT_e$ increases (see bottom panel of Fig. ~\ref{lc_1999}).
% which indicate on significant inferaction between 
%the TL and the NS surface emission. 
We should emphasize once again  that 
%a reduction of 
the photon index, $\Gamma_2$  correlates with   $kT_{s2}$ (see Fig. \ref{two_state_spectra}, left panel) 
which clearly indicates  the CC expansion. The disk seed  photons, in this case, come from the cooler outer regions [see  \cite{STSS15}  for a possible geometry of the source].
%  during this transition phase.

%((((
\section{Discussion \label{disc}}

\subsection{Quasi-constancy of the index is a  signature of a NS  \label{constancy}}

Using our analysis of the index evolution  in 4U 1700-37
we have firmly established  the { $\Gamma_1$  quasi-constancy versus 
Comptb normalization {\bf  $N_{Com1}$,  (proportional
% to the (disk) mass accretion rate, 
to $\dot M$})} and also versus other parameters, $kT_{s}$ and $kT_e$ 
(see Fig. \ref{two_state_spectra}).   FT10 and ST11 argue that this  $\Gamma$  stability is 
a signature  of solid surface of a compact object, {\bf pointing to a NS presence} in the HMXB 4U~1700-37. 
On the other hand the photon index $\Gamma_2$ decrease   is directly related to  expansion of {\bf the transition layer (TL)} due to high radiation pressure  when  the index  drops  and    the plasma temperature $kT_e$ increases (see Fig. \ref{two_state_spectra}).
TSS14 explained this observational effect. {\bf  At high  luminosity  the radiation pressure from the NS photons  stops the falling plasma. As a consequence   the gravitational energy deposition  only takes place in an outer part of the transition layer where the plasma temperature is high  (above  30 keV)}. At such  temperatures  the critical luminosity is not achieved.      
% which indicate on significant 
%inferaction between the TL and the NS surface emission. 
Observationally, the expansion of the TL is also confirmed by a decrease of the seed photon temperature
 $kT_{s2}$.  
%In addition, the reduction of the photon index $\Gamma_2$ phase is accompanied  with the 
%$kT_s2$ decreesing, which clear indicate on the cooling (and possibly, expansion) of the 
%TL during this transition phase. 
Thus, the photon index $\Gamma_2$ demonstrates the 
stability at the sub-Eddington luminosity and a decrease %of the Photon Index $\Gamma_2$ 
at the high plasma temperature state as was previously found in Sco~X-1 (TSS14) and 4U1705-44 (STSS15). 

Note that 
%NEW
{ 
only three sources, 4U~1705-44, 4U~1700-37  and Sco X-1 show a drop of the
photon index (see Figure \ref{gam_te_7obj}).
% at high electron temperatures. 
However, 
}
%END NEW
this index effect  in 4U~1700-37  begins at lower  $kT_e\sim 40$ keV, 
%the decreasing index phase  of 4U~1705-44  begins at higher electron temperature ($kT_e\sim 80$ keV), 
than that  in Sco~X-1 ($kT_e\sim 60$ keV) and 4U~1705-44 ($kT_e\sim 80$ keV). We should point out that in Sco~X-1  the maximum of the plasma temperature, $kT^{max}_e$ exceeds 100 keV while  that for 4U~1700-37 and 4U~1705-44 are in the range $kT^{max}_e\sim 95-100$ keV
%which is possibly related 
%to different X-ray luminosity levels 
(see Figure \ref{gam_te_7obj}). % (for  $atolls$ and {\it Z-}sources). 
%However, 
% 4U~1705-44 emits  very close to a {\it Z-}source luminosity range as can be seen  from Fig.~\ref{T_e vs lum_5obj}. %, \ref{HID_7object}.

\subsection{Possible detection of the switches between the wind and disk accretion regimes}
%It is worthy to note that 
The optical component in 4U~1700-37 is characterized by a strong  wind as it is usually observed  in  O$-$supergiants. The wind velocity near inner Lagrange point increases (Abubekerov, 2004). SgXB 4U~1700-37 have also been proposed to undergo episodes of disk accretion 
due to the formation of short lived accretion disks  
%especially close to the periastron passage in eccentric systems 
(see, e.g., Fryxell \& Taam 1988). %; Taam \& Fryxell 1988, ApJ, 327, L73; 
%Taam et al. 1988, ApJ, 331, L117). 
Generally, the angular momentum of the fast wind launched from a supergiant star is  too low to form a disk (Ducci et al. 2009).
However, this is not true if the velocity of the wind is strongly reduced by the effect of ionization of the X-rays emitted from the NS (in this case the line driven acceleration mechanism can be nearly turned off). As discussed recently by Romano et al. (2015), in systems with an orbital period of a few days (5 -- 10 days) and a moderate eccentricity, a disk is likely to form when the velocity of the wind from the massive companion drops by a factor of $\sim$100 (i.e. from thousands to tens of km/s). Thus, when accretion switches from the wind to the disk mode,  X-ray  luminosity  is expected to rise significantly ($\simeq 10^{37}-10^{38}$ erg/s), due to the enhanced  disk mass accretion rate.  
%and  
%that can be achieved through 
%the presence of the disk. 
A rapid spin up phase is also expected to take place due to the
angular momentum of the disk material  that acquired by the NS during the accretion process (Klochkov et al. 2011 and
% A\&A, 536, 8; 
Jenke et al. 2012). 
%The rapid switch between disk and wind accretion is an extremely interesting and poorly understood phenomenon, as it happens practically only sparse and low resolution X-ray data co be collected so far during such transitions (transitions are expected to be rapid - thousands of seconds - and sporadic, so diffcult to catch. 

Variable inner absorption is an  observational evidence in favor of accretion disk formation in 4U~1700-37. 
In fact, the absorption amount  should rises as the compact object appears behind the wind  dense parts,  close to  eclipse ingress and egress. However,  this absorption is asymmetric about orbital phase $\varphi=0.5$ in this system (Branduardi, Mason, \& Sanford 1978).  Mason et al. 1976 and  Haberl et al. 1989  that a sharp increase in absorption at  $\varphi\ge 0.6$ is related to   a region of  increased density. 
%trailing the compact object in its orbit, 
%such  a photo-ionisation wake (Blondin et al. 1990). 
%The presence of  a wake in 4U~1700-37 has been confirmed with optical spectroscopy (Kaper et al. 1994) at orbital  phases around 0.6. 
Boroson et al. (2003) detected a ''soft excess'' in the flux at $\varphi\simeq$ 0.7.  Haberl, White, \& Kallman (1989) pointed out  that this  absorption excess at $\varphi > 0.6$ could be due to 
an accretion   taking  8\% of the gas flow that accretes through 
%onto the compact object via 
the disk.

Using {\it Beppo}SAX, {\it Suzaku}  and {\it RXTE} observations, 
we find  two %three 
blackbody emission sources: 
the first one is probably  associated  with the disk and the second  is presumably associated with the neutron star (NS), 
% and the third %another one is related to the CC region, 
for which soft photon temperatures are about %0.7 keV 
1 keV  or less and 1.4 keV, respectively.
We also established  that some spectra {\bf can be fitted with an  additional 
 %three abovementioned sources of 
{Blackbody}  (with  normalization $N_{bb}$ and $kT_{bb}\sim 0.6$ keV), while  other  spectra  do not  require this component.
%   {Blackbody}.
% emmision associated with the accretion disk 
%(more properly, with very low normalization $N_{BB}$). 
Note, the absense of the Bbody at 0.6 keV is generally 
related to the low/hard states of 4U~1700-37, while its detection  is strictly accompanied by flaring events. These episodes are marked by vertical strips in Figure~\ref{lc_1999} to relate these episodes to transient accretion disk events}.  One can also  see 
that these flaring events  are associated with drops of  the  index $\Gamma_2$  and surges  
of $N_{bb}$  ($\propto \dot M$)  by factor 10. Therefore, we can relate these moments with 
 {transient} accretion disk events. In fact,  we have  found  six  of these intervals, in which we suggest  switches between the wind  and disk accretion regimes.

In the case of very low $N_{bb}$ at 0.6 keV we fix  $N_{bb}$ at 0.001$\times L_{39}/D^2_{10}$, {\bf  in order to model and compare  our results for all spectral states    
and  use} the same model for all states.  This procedure allows us to take into account a weak accretion disk.

%In the end of this discussion section we should point out  a few of interesting questions and comments  which can be raised by a potential reader. For example,  one may put a fair question on a physical sense of  our spectral model, which we use  to fit the data, and for the final  intepretation.
% which we use  to fit the data.
% but has a difficult physical
%> intepretation.
% For example, if  our spectral model  is unique  one can simply use a PL (power-law model) instead of the second COMPTB.
%In fact,  our spectral model is unique  {\bf because we assume   two
%sources} of soft photons.
%The first one is the NS surface and another is  an extended configuration which can be an accretion disk. As for  a  simple power-law,  as a  component of the spectral model then  a natural question is arised  what is an origin this power law (PL) how  this PL  is formed. 
 In our case, the extended power-law is a tail of the Comptonization spectrum which is created in the outer hot part of the Compton cloud and  we have already demonstrated and explained this PL phenomenon for a number of NS sources (see  STSS15, TSS14 for details).  Thus, our spectral  model has a basic, physical meaning, because it relies on the Comptonization of two sources of soft photons, {\bf one coming from the NS and the other has  the disk origin}. As for  the Bbody source,  these    soft photons  come from outer part of the extended disk.

{\bf It is worth noting that we keep $\log A=2$ when $A\gg1$ because   
%In fact, when $A$ is much %greater than one then 
an  illumination factor f=A/(1+A) is close to one in this case. 
We should also clarify a relation  between  the XSPEC COMPTT and COMPTB models because 
they are widely used for modeling
%many people of our community use these models 
 X-ray data.   
The COMPTT model  parameters are the optical depth, which value depends on an assumed geometry, and the plasma temperature $T_e$.
We prefer  to be independ on any assumption  of the plasma cloud (TL) geometry. Thus,   we use the COMPTB model where the main
parameters  are the spectral index $\alpha$ and  the plasma temperature $T_e$. 
% there is any degenerac y in the parameters (as $\tau$ and $T_e$ in COMPTT) and 
%thus our Figure \ref{two_state_spectra} come from a degeneracy among the two COMPTB components. In fact,  we do not use  COMPTT model, instead we use COMPTB where the main
%parameters of the model are the spectral index $\alpha$ and  the plasma temperature. 
Instead of  using  the optical depth and plasma temperature provided by 
%to obtain the spectral index as that in the case of 
the COMPTT  one can simply apply the COMPTB model}.

Our  Figure \ref{two_state_spectra} presents the main result of the paper where we show that  the photon index $\Gamma$ ($\alpha$+1) is   around 2 while the CC plasma temperature is less than 30 keV. Note, that the applied electron cross-section is Thomson  in this case.  However, when the CC plasma temperature increases above 30 keV the electron cross-section drops  and  thus, the critical  luminosity increases. In this case,  $kT_e>30$ keV,  
the emergent luminosity is less than the critical  and thus, the radiation pressure originated at the NS surface cannot stop the accretion flow. 

Note,  4U 1700-37 HID is similar to those of LMXBs while the CCD is different. 
%There is no hint of  atoll or   Z-source in the CCD. 
Thus, we can conclude   that 4U 1700-37  behaves in the same way as a NS in LMXB in terms of physical quantities (the spectral index and the CC plasma  temperature)  keeping, however in mind that CCD is only  a qualitative description of the data.
Indeed,  our Figure \ref{CCD_HID_1700}  presents clearly  the spectral softening when the count rate increases.  

One possible issue  for the presented study 
%can also state that  the main problem of our study 
is that the magnetic field which can be as high as normal X-ray pulsar, see aforementioned claims using the data analysis of the appropriate {\it Beppo}SAX observations
[Reynolds et al. (1999) and Jaisawal \& Naik, 2015].  The absence of  pulsations from this source %and people even think
even leads to a suggestion  that  the source could be a BH source.   In fact, we have  found using our analysis that the illumination factor of the Compton cloud,  $f$ is very close to 1 (or $\log A\gg1$) in most of the  observations. This means that the NS is embedded in a thick
Compton cloud that does not allow a direct view of the NS surface, making
difficult to estimate  the NS magnetic field.
% and thus, one can conclude that the NS is embedded in a thick Compton cloud. In this case, one cannot see  a direct radiation  from the NS surface and the NS magnetic field is not possible to estimate.     

\section{Conclusions \label{summary}} 

Applying {\it Suzaku},  {\it Beppo}SAX,  and {\it RXTE} observations we studied the correlations of spectral %, timing 
properties, with mass accretion rate, $\dot M$ observed in the non-pulsating high mass binary, 4U 1700-37.
%X-ray source 4U~1700-37 %Sco~X-1  

 We find  that  all broad-band energy spectra of this source 
%during all %these 
%states %transitions %spectral states 
can be adequately presented as a sum  of
%of  low-temperature $Blackbody$ component, 
 two Comptonization %({\it COMPTB}) 
spectra associated  with the  {seed} photon temperatures, $T_{s1} \sim 1.3-2$ keV  and 
$T_{s2} =0.8-1.1$ keV.
We  also  put  an iron-line component %and a cyclotron absorption line component ($cyclabs$) 
to the model. 
% which leads to improved fits. 
 
\begin{figure}
% \centering
\includegraphics[scale=0.95,angle=0]{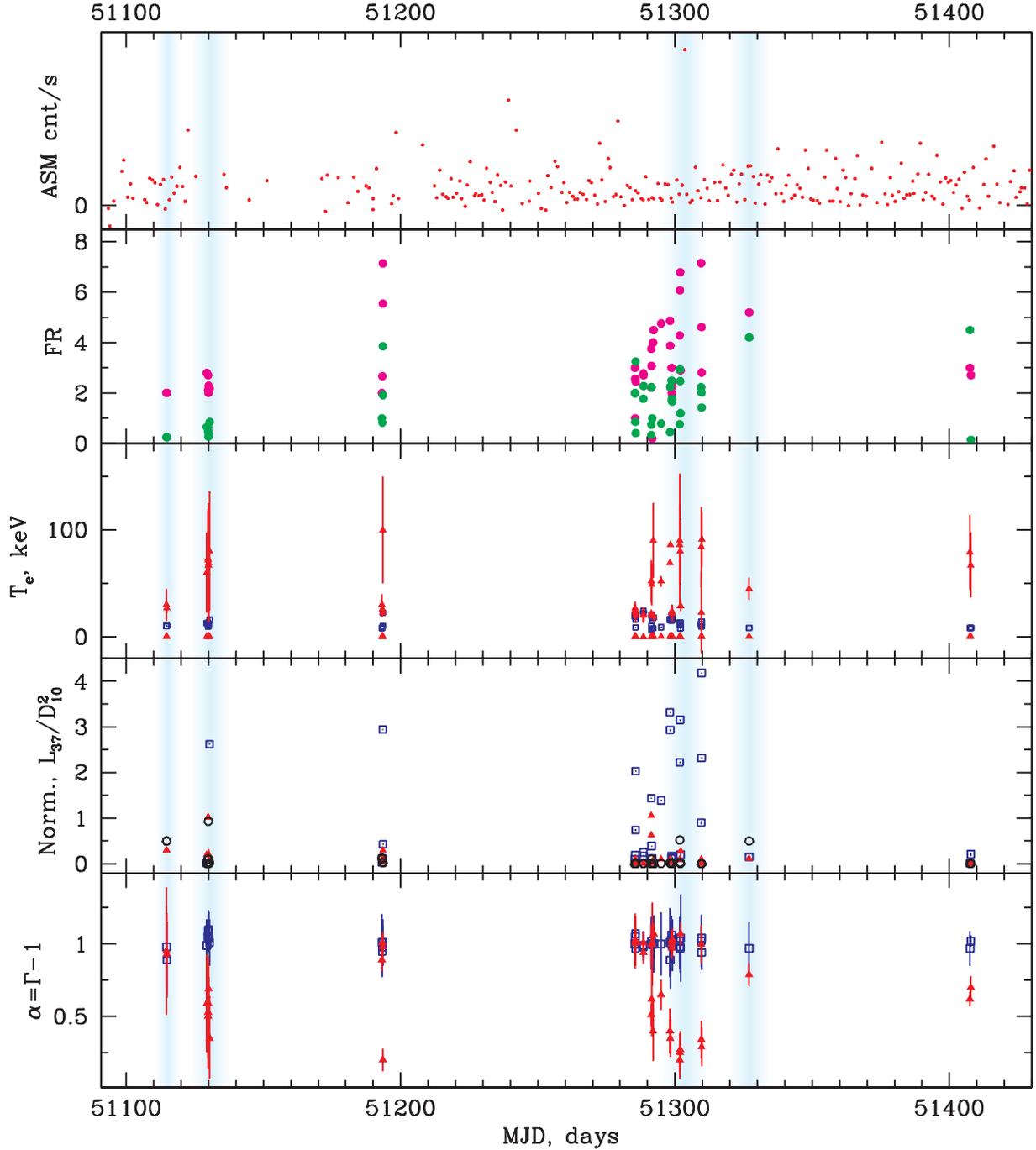}    
      \caption{
From top to bottom: evolution of {\it RXTE}/ASM count rate, 
%{\it From Top to Bottom:}
%Evolutions of  count rate [2-9 keV] in counts s$^{-1}$ with 16~s time resolution, 
 {Ratios} of the (10-50 keV)/(3-10 keV) fluxes ($pink$) and the (50-200 keV)/(3-10 keV) fluxes (green), 
%the  model flux in the 3-10 keV and 10-50 keV energy ranges ($black$ and $green$ points, respectively),  
 $kT^{(1)}_e$ (blue) and $kT^{(2)}_e$ (red) 
%of the $Comptb1$ and $Comptb2$ components, respectively, 
in keV,  
blackbody normalizations of Comptb1, Comptb2 and Bbody components (red, blue and black respectively),    
and  the spectral indices $\alpha_1$ and $\alpha_2$ 
%($\alpha_{1,2}=\Gamma_{1,2}-1$) 
(blue and red) for Comptb1 and Comptb2  components, respectively for  1999 evolution %transition 
events (R2 set). 
The %dipping/flaring 
phases of the light curve, related 
%(based on timing analysis) 
%to the {\it reduced spectral index $\alpha_2$}, %flaring branch}, 
to the {increased Bbody component}, %flaring branch}, 
are marked with blue vertical strips. 
}
\label{lc_1999}
\end{figure}

\begin{figure}[ptbptbptb]
    \includegraphics[width=17cm,clip]{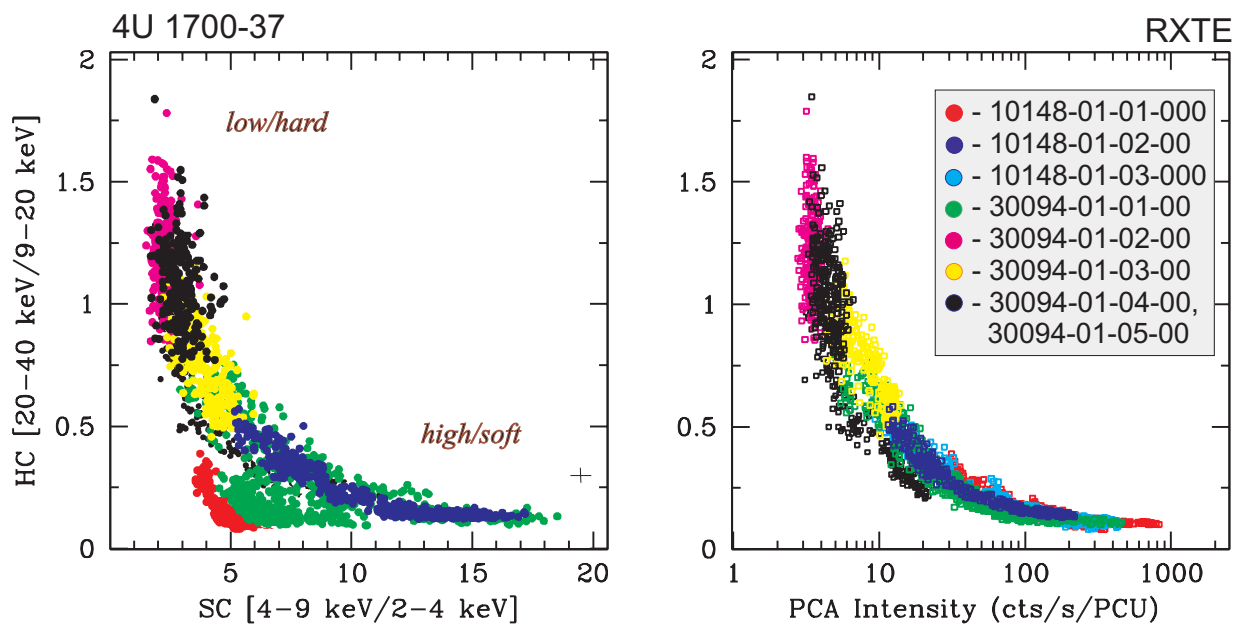}
%68 446 513 774
      \caption{
CCDs ({left} panel) and HIDs ({right} panel) for observations of 4U~1700-37 used in our analysis, with  
%The bin size is 16 seconds. 
bin size 16 s. 
%The typical error bars are smaller than the symbol size. 
 The typical error bars for the colors are shown in the right  bottom  corner of the left panel while  errors of  the intensity are negligible.  
The sets are indicated by different colors: red (ObsId 10148-01-01-000), blue (ObsId 10148-01-02-00), 
bright~blue (ObsId 10148-01-03-000), green (ObsI 30094-01-01-00), crimson (ObsId 30094-01-02-00), 
yellow (ObsId 30094-01-03-00) and black (ObsIds 30094-01-04-00, 30094-01-05-00).
}
      \label{CCD_HID_1700}
 \end{figure}

\begin{figure}[ptbptbptb]
%  \begin{figure}
 %\centering
%\includegraphics[scale=1.0,angle=0]{f2_1700.eps}
\includegraphics[scale=1.0,angle=0]{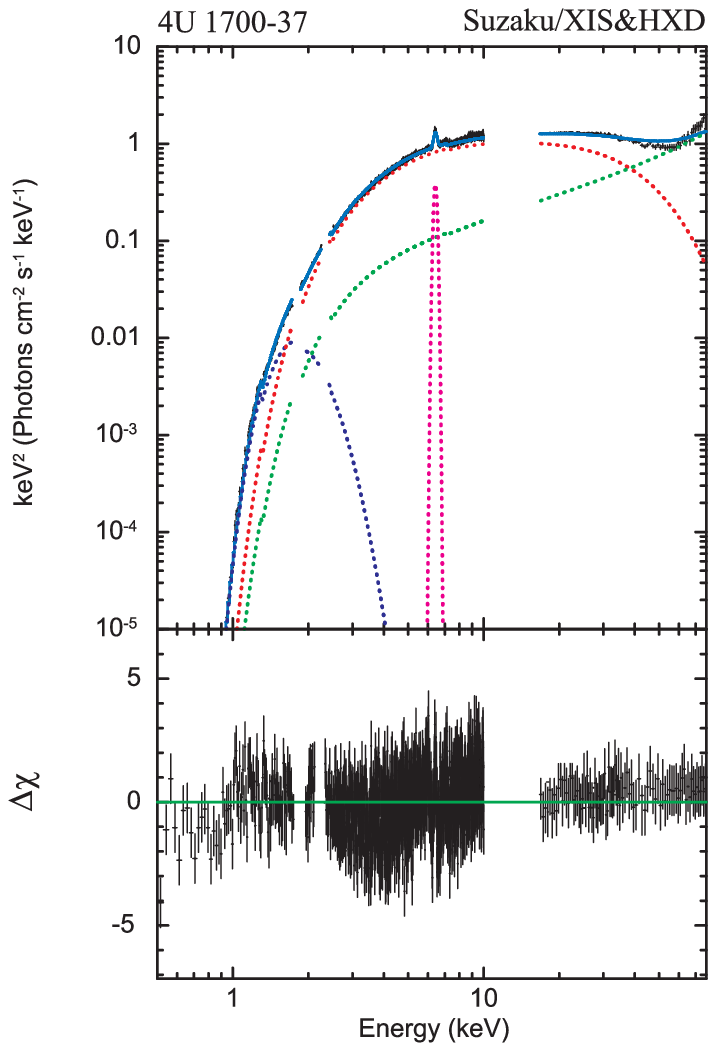}
\includegraphics[scale=1.0,angle=0]{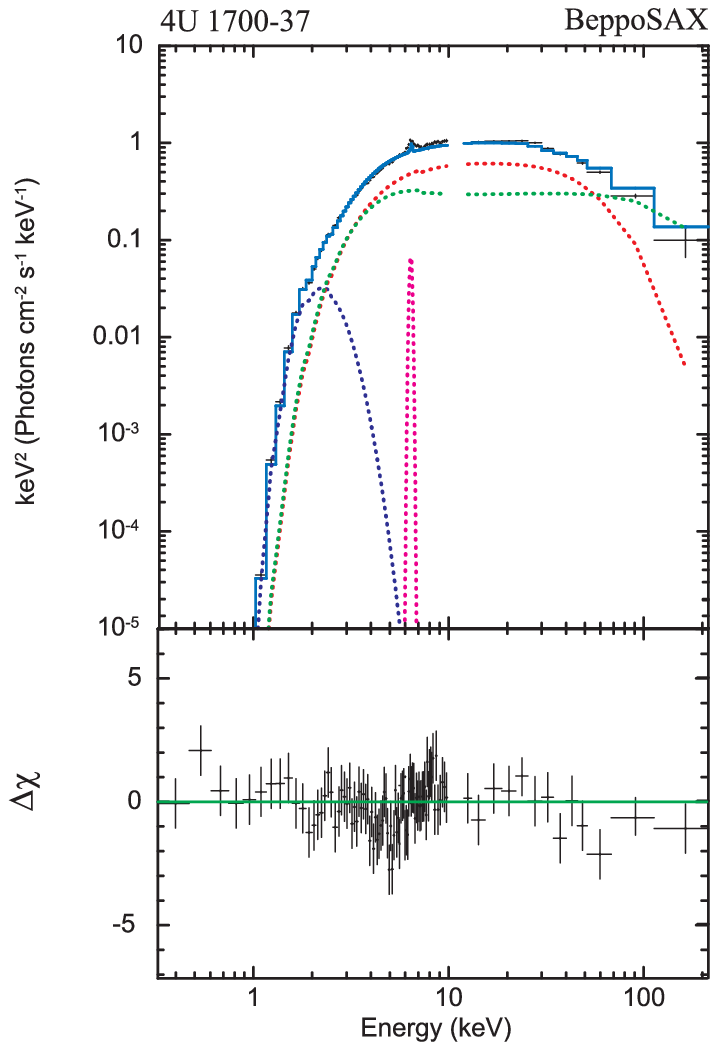}
      \caption{Energy spectrum of 4U~1700-37 obtained using %the XISs and PIN detectors of the 
{\it Suzaku} ({left} panel) and {\it Beppo}SAX ({right} panel) observations along with
the best-fit model phabs*(bbody + comptb + comptb + gaussian). 
%comprising a partial covering NPEX contin-
%uum model, three Gaussian functions for emission lines and a
%cyclotron absorption component.
}
\label{spectrum_Suzaku_SAX}
%\end{figure}
\end{figure}

\begin{figure}
%   \resizebox{\hsize}{!}{\includegraphics{f3_1700.eps}}
   \resizebox{\hsize}{!}{\includegraphics{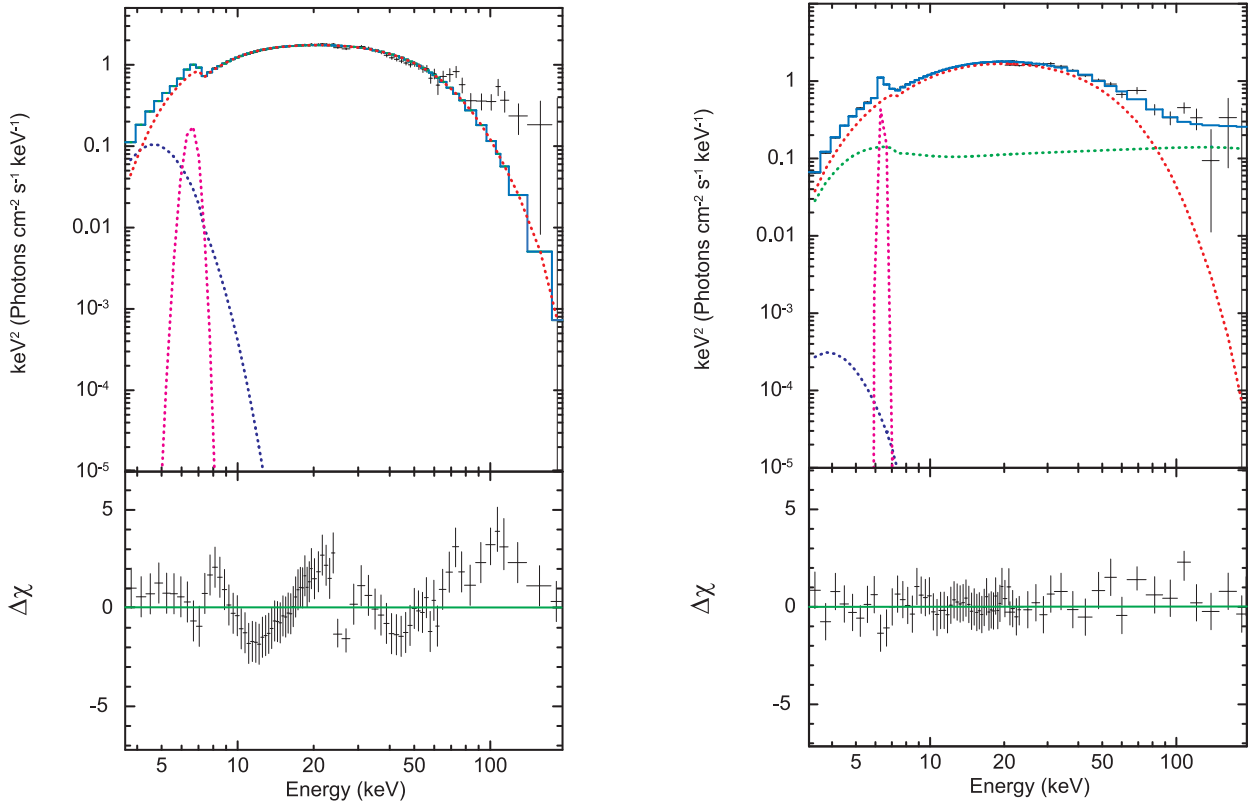}}
   \caption{On the spectral model choice for 4U~1700-37 using {\it RXTE} observation 
%On the choise of the spectral model for 4U~1700-37 by the example of RTXE observation 
(ObsId=10148-01-02-00; 1996, September 13).
% Representative spectrum of 4U~1700-37 during $high/soft$ events on 1996, September 13 (observation 10148-01-02-00) 
%in $E*F(E)$ units modelled by two different models.
Left: the spectrum along with the fit residuals %$\Delta\chi$ 
for the model fit, phabs*(bbody+comptb+gauss), which includes a 
single Comptb component ($\chi^2_{red} = 12.3$ for 86 dof), and right: the best-fit spectrum and $\Delta\chi$ for the model fit, 
phabs*(bbody+comptb1+\-comptb2+gauss), which consists of two Comptb components ($\chi^2_{red} = 1.12$ for 81 dof).
The data  are shown by crosses and the best-fit spectral  model %  {\it wabs*(blackbody+Comptb1+Comptb2+Gaussian)} 
is presented by light-blue line. The model components  are indicated by dark-blue, red, green 
and crimson lines for { Blackbody}, {Comptb1}, {Comptb2}  and  {Gaussian} components, respectively. 
% On the choise of the spectral model for 4U~1700-37 by the example of 10148-01-02-00 observation (1996, September 13).
}
   \label{rxte_spectra_one_two_comptb}
 \end{figure}

 {\bf Our  model of  the 4U~1700-37 spectra  allows us to separate the contributions of  
two specific  spectra,  related to %both 
the hard and soft X-ray components 
%into common mass accretion rate $\dot M$ 
along the CCD and HID 
 presumably  dictated by  $\dot M$.  %In fact,  
%mass accretion rate is proportional to 
The observed  luminosity  determined by
 two  {normalizations}  $N_{Com1}$ and  $N_{Com2}$ of the  Comptonized components %(see Eq.~1) % \ref{comptb_norm}) 
which are proportional to $\dot M$.  % $N_{Com}$. 
%mass accretion rate.    
%From one hand, obtained in this Paper 
We also found that  the soft Comptonization
component (associated with the NS seed photons) dominates over the 
hard Comptonization component (associated with  the   
disk seed photons). This effect is 
clearly visible in the second panel from the bottom in Figure~\ref{lc_1999}}.

%  (see the second panel from
%the bottom of Figure~\ref{lc_1999}).

During X-ray flares, in a surge of
 $\dot M$,   we observe an increase of  $N_{Com2}$ %decreasing of $N_{Com1}$ % and decreasing of $N_{Com2}$ 
%from  the $island$ to $banana$ states %$soft~apex$ to the $hard~apex$  
%in CCD %HFD 
 %HID_Sco})  %. and from $soft~apex$ to
 correlated with a rise of the plasma temperature $kT_e^{(2)}$  (see 2nd and 3rd  panels from the bottom of  Figure \ref{lc_1999}).
%  indicates to a decrease of . %from the {\it lower} normal branch  to 
%the {\it upper} normal branch.
% The normalization parameter $N_{Com2}$ is related to the soft NS Comptonized component,  %surface of Sco~X-1 
%and it is related  to a total mass accretion rate $\dot M_{tot}$.  %One can see from Fig. \ref{te1-Zstate}, panel b there,  
%how the total mass accretion rate, or  $N_{Com2}$, increases when  $kT_e$ increases and using Fig. \ref{HID_Sco} one can also see how the hardness ratio increases from the  {\it lower} FB  to the {\it upper} FB.
%We should also point out that a comparative analysis using our model shows that various NS LMXBs are also  well traced by soft   luminosity in the energy range from 10 to 50 keV.

Our spectral analysis of 4U 1700-37
%using our model 
also  discovers 
the stability of photon indices $\Gamma_1$ and $\Gamma_2$ 
around 2 during the %$island$ and $banana$ % HB/NB/botFB 
states which relate to  the CC plasma temperature range of   $kT^{(2)}_e$ 
 from 2 keV to  30 keV, while a drop of 
$\Gamma_2$ is seen  during a  rise of  $kT^{(2)}_e>$ 40 keV (see Figure \ref{two_state_spectra}).  %the mid-topFB. 
We interpret the quasi-stability  of $\Gamma$
%detected quasi-stability 
%of the photon indices of the Comptonized components around 
around  
 2 
%during the low tempwrature phase %HB -- NB -- botFB %{\it Comptb1/Comptb2})
in the framework of a model in which the gravitational energy release  occurs in the transition layer  (TL) and  this energy deposition in the TL   greatly exceeds   that in the  disk.
% The stability of the index at a value of approximately 2   
%index stability phase %effect 
%is now established for the Comptonized spectral components of 4U~1700-37. %Sco~X-1 over %through the HB -- NB -- botFB. 
%when $kT^{(2)}_e>$ 85 keV. 
The similar  effect  was previously  revealed in LMXBs: 
  atolls,  
4U~1705-44, 4U~1728-34, GX~3+1, 4U~1820-30 and  {Z-}sources, Sco~X-1, GX~340+0  
%through all %spectral states 
%and { Z-}source Sco~X-1  during HB-NB-botFB 
(see Figure \ref{gam_te_7obj}).   In addition to the  index plateau phase  
of $\Gamma_{1,2}\sim2$   we also reveal the index decrease   for the hard Comptonization component { detected over the flaring branch (FB)}, in addition to the index plateau stage of $\Gamma\sim2$. 
%in 4U~1700-37 %Sco~X-1 
 Note, that a  similar phase is observed  in 4U~1705-44 over the upper banana branch and  in Sco~X-1 at the FB  which is an additional argument  for the NS presense  in 4U 1700-37.
 
We can  interpret this index decrease using 
 the  model in which the gravitational energy release occurs only in an outer part 
of the Compton cloud (CC) (see details in TSS14).   The radiation force acting on the accretion  flow stops it  near the inner boundary  of the CC  where $T_e$ is actually dictated by the photon flux from the NS surface and as a result   the Klien-Nishina cross-section  rises  with a  plasma temperature decrease  
% inward stops the accretion flow 
(see \S 4 of TSS14 for details). %\ref{index reduction}).  

%During this decreasing index stage 
%index reduction stage, 
%the  X-ray spectra of both 4U~1700-37, 4U~1705-44 and Sco~X-1 exhibit  an increase of the  electron temperature of the hard Comptonized component %$T_e^{(1)}$ 
%from  40 to  180 keV. 
%which indicate on similarity of $atoll$-source 4U~1705-44 and 
%Z-source Sco~X-1.   
Note,  in BHs the photon  index rises 
and then saturates when the luminosity increases (see e.g. ST09).  This   index behavior in a BH  is in contrast  to that seen in 4U~1700-37, 4U~1705 and Sco X-1. 

%{\it Beppo}SAX and  {\it Suzaku}  observations provide us a hint of a switch between the wind and disk accretion in HMXB 4U~1700-37.  
In our analysis  we establish the robust nature of the parameter correlations in  4U~1700-37 and we  want to point out that  this source  is a neutron star one.  
% resulting from the application of our model 
%points to fundamental insights into the complex, and poorly understood, behavior of HMXBs.

\begin{acknowledgements}
We appreciate editing the text  of the paper by Demos Kazanas and his valuable suggestions which improve the paper presentation.  We should  acknowledge  hot discussion  of the paper results  with Sergio Campana (see  Discussion section for details).
We also want to acknowledge the referee’s efforts on the clear presentation of our paper.
\end{acknowledgements}

%888

\newpage

%
% Table 1
%
%
%  TABLE 1 - BINARY PARAMETERS
%

\begin{deluxetable}{llc}
%\begin{deluxetable}{ll}
%%%%%\rotate
\tablewidth{0in}
\tabletypesize{\scriptsize}
    \tablecaption{Binary parameters of the 4U 1700-37 system.} % (from literature).}\vspace{1em}
%    \tablecaption{Parameters of the 4U~1700-37.} % (from literature).}\vspace{1em}
    \renewcommand{\arraystretch}{1.2}
\tablehead{  
Parameters               & Value & References}
%Parameters               & Value }
\startdata
  Orbital period, $P_{orb}$       &   3.412 days                              & (1) \\
  Orbital eccentricity, $e$       &   $\le$ 0.01                              & (2) \\
  Orbital inclination, $i$        &   66$^\circ$                              & (2) \\
  Mass of O6f star, $M_{V}$       &   27.4, $\sim 30$ $M_{\odot}$             & (2, 3)\\
  Mass of compact object, $M_{X}$ &   2.1 - 2.3, $\sim 2.6$ $M_{\odot}$       & (2, 3) \\
  Source distance, $D$            &   1.9 kpc                                 & (4) \\
%
%  Orbital period, $P_{orb}$       &   3.412$^1$ days  \\
%  Orbital eccentricity, $e$       &   $\le$ 0.01$^2$  \\
%  Orbital inclination, $i$        &   66$^\circ$ $^2$ \\
%  Mass of O6f star, $M_{V}$       &   27.4$^3$, $\sim 30^2$ $M_{\odot}$ \\
%  Mass of compact object, $M_{X}$ &   2.1 - 2.3$^3$, $\sim 2.6^2$ $M_{\odot}$  \\
%  Source distance, $D$            &   1.9$^4$ kpc   \\
      \enddata
    \label{tab:table_param}
References:  
(1) Boroson et al. 2003; Haberl, White \& Kallman 1989;  %UHURU 2011; 
(2) Rubin et al. 1996; 
(3) Abubekerov 2004; %, 2005;
(4) Ankay et al. 2001.
\end{deluxetable}

\newpage
 
% &&&&& Table 2

% TABLE 2 - SAX and Suzaku DATA  &&&&&&&&&&&&&& ApJ format &&&&&&&&&&&&&&&&&&&&&&&&&&&&&
\begin{deluxetable}{l l l l l l c}
%%%%%\rotate
\tablewidth{0in}
\tabletypesize{\scriptsize}
    \tablecaption{The list of $Beppo$SAX and $Suzaku$ observations of 4U~1700-37 used in our analysis.} 
    \renewcommand{\arraystretch}{1.2}
%\tablehead{Satellite &  Obs. ID        & Start time (UT)  && End time (UT) & MJD interval}
\tablehead{Satellite &  Obs. ID        & Start time (UT)  & & End time (UT) & MJD interval & Exposure (ks)}
\startdata
$Beppo$SAX & 20339001$^1$       & 1997 April 1  11:20:47 & & 1997 April 1 21:51:19 & 50,539.4 -- 50,539.9 & 23.7$^a$\\
$Suzaku$   & 401058010$^2$      & 2006 Sept. 13 10:44:38 & & 2006 Sept. 14 22:03:14 & 53,991.4 -- 53,993.1 & 82.1$^b$\\
%
%$Beppo$SAX & 20339001$^1$       & 1997 April 1  11:20:47 & & 1997 April 1 21:51:19 & 50,539.4 -- 50,539.9 &\\
%$Suzaku$   & 401058010$^2$      & 2006 Sept. 13 10:44:38 & & 2006 Sept. 14 22:03:14 & 53,991.4 -- 53,993.1 &\\
      \enddata
   \label{tab_sax_suzaku}
References. 
(1) \cite{Reynolds99}; 
(2) \cite{Jaisawal_Naik15}. 
%Piraino et al., (2000)
$^a$ effective exposure using MECS instrument, while those  in  LECS  and PDS instruments are 12.2 ks  
and 10.7 ks, respectively;
%is indicated The $Beppo$SAX exposures in the LECS, MECS, HPGSPC, and PDS instruments are 12.2 ks, 23.7 ks, 11.2 ks, and 10.7 ks, respectively.
$^b$  effective exposure using HXD instrument while that   in XIS instrument is 81.5 ks.
%The Suzaku observation was performed in "XIS nominal" position with an effective exposure of 81.5 ks and 82.1 ks for XIS and HXD, respectively.

\end{deluxetable}
%\begin{deluxetable}{l l l l l l c}
%%%%%\rotate
%\tablewidth{0in}
%\tabletypesize{\scriptsize}
%    \tablecaption{The list of $Beppo$SAX and $Suzaku$ observations of 4U~1700-37 used in our analysis.} 
%    \renewcommand{\arraystretch}{1.2}
%\tablehead{Satellite &  Obs. ID        & Start time (UT)  && End time (UT) & MJD interval}
%\startdata
%$Beppo$SAX & 20339001$^1$       & 1997 April 1  11:20:47 & & 1997 April 1 21:51:19 & 50,539.4 -- 50,539.9 &\\
%$Suzaku$   & 401058010$^2$      & 2006 Sept. 13 10:44:38 & & 2006 Sept. 14 22:03:14 & 53,991.4 -- 53,993.1 &\\
%      \enddata
%   \label{tab_sax_suzaku}
%References. 
%(1) \cite{Reynolds99}; 
%(2) \cite{Jaisawal_Naik15}
%Piraino et al., (2000)
%\end{deluxetable}
%'
\newpage

% &&&&&& Table 3  &&&&&&&&&&&&&& ApJ format &&&&&&&&&&&&&&&&&&&&&&&&&&&&&

% TABLE 3 - RXTE log

\begin{deluxetable}{l c r c c c}
%\begin{deluxetable}{l c r c c}
%\begin{deluxetable}{l c c c}
%%%%%\rotate
\tablewidth{0in}
\tabletypesize{\scriptsize}
    \tablecaption{The list of {\it RXTE} observations of 4U~1700-37.}
    \renewcommand{\arraystretch}{1.2}
\tablehead{Number of set  & Dates, MJD   & Exposures, s      & & {\it RXTE} Proposal ID&  Dates UT}
%\tablehead{Number of set  & Dates, MJD          & RXTE Proposal ID&  Dates UT}
\startdata
R1  	       &    50,334-50,339      & 29~685     & & 10148$^1$     & Sept. 8 -- 13, 1996          \\
R2             &    51,285-51,408      & 2~687~688  & & 30094$^1$     & April 17 -- Aug. 17, 1999    \\
R3             &    51,129-51,193      & 53~238     & & 30095$^{1}$   & Nov. 12 1998 -- Jan. 15, 1999\\
R4             &    51,776-51,779      & 1~945~005  & & 40068$^{1}$     & Aug. 20 -- 23, 2000          \\
R5             &    52,890.09-52,890.2 & 9~535      & & 80105$^1$     & Sept. 8, 02:19:28 -- 03:51:56, 2003    \\
%
%R1  	       &    50,334-50,339      & 29,7   & 10148$^1$     & Sept. 8 -- 13, 1996          \\
%R2             &    51,285-51,408      & 2687.7 & 30094$^1$     & April 17 -- Aug. 17, 1999    \\
%R3             &    51,129-51,193      & 53,2   & 30095$^{1}$   & Nov. 12 1998 -- Jan. 15, 1999\\
%R4             &    51,776-51,779      & 1945,0 & 40068$^{1, 2}$     & Aug. 20 -- 23, 2000          \\
%R5             &    52,890.09-52,890.2 & 9,5    & 80105$^1$     & Sept. 8, 02:19:28 -- 03:51:56, 2003    \\
      \enddata
      \label{tab:list_RXTE}
Reference: 
%References: 
(1) Dolan 2011 %; 
%(2) Kong \& Di Stefano 2005; 
%(2) Boroson et al. 2003
\end{deluxetable}

%\begin{deluxetable}{l c c c}
%%%%%\rotate
%\tablewidth{0in}
%\tabletypesize{\scriptsize}
%   \tablecaption{The list of {\it RXTE} observations of 4U~1700-37.}
%    \renewcommand{\arraystretch}{1.2}
%\tablehead{Number of set  & Dates, MJD          & RXTE Proposal ID&  Dates UT}
%\startdata
%R1  	       &    50,334-50,339      & 10148$^1$     & Sept. 8 -- 13, 1996          \\
%R2             &    51,285-51,408      & 30094$^1$     & April 17 -- Aug. 17, 1999    \\
%R3             &    51,129-51,193      & 30095$^{1}$   & Nov. 12 1998 -- Jan. 15, 1999\\
%R4             &    51,776-51,779      & 40068$^{1, 2}$     & Aug. 20 -- 23, 2000          \\
%R5             &    52,890.09-52,890.2 & 80105$^1$     & Sept. 8, 02:19:28 -- 03:51:56, 2003    \\
%      \enddata
%      \label{tab:list_RXTE}
%References: 
%(1) Dolan 2011; 
%(2) Kong \& Di Stefano 2005; 
%(2) Boroson et al. 2003
%\end{deluxetable}

\newpage

%&&&&& Table 4  &&&&&&&&&&&&&& ApJ format &&&&&&&&&&&&&&&&&&&&&&&&&&&&&

%
% TABLE 4 SAX Suzaku  ANALYSIS
%

\begin{deluxetable}{lllll}
%%%%%\rotate
\tablewidth{0in}
\tabletypesize{\scriptsize}
    \tablecaption{Best-fit parameters of spectral analysis of $Beppo$SAX (ObsId=20339001), $Suzaku$ (ObsId=401058010) and 
$RXTE$ (ObsId=10148-01-02-00) observations of 4U~1700-37$^{\dagger}$  
%in 0.3 -- 150~keV  energy range
using  different models. 
%Parameter errors correspond to 90\% confidence level.
}
% \label{tab:fit_table_SAX_Suzaku}
    \renewcommand{\arraystretch}{1.2}
\tablehead{Model & Parameter & 00-20339001  & 0-401058010 & 10148-01-02-00}
\startdata
phabs    & N$_H$ (cm$^{-2}$) &  2.8$\pm$0.1    & 3.78$\pm$0.02    & 5.7$^{frozen}$\\ 
bbody    & kT$_{BB}$ (keV)   & 0.99$\pm$0.02   & 0.98$\pm$0.03  & 0.48$\pm$0.06 \\
         & N$_{BB}^{\dagger\dagger}$ & 0.07$\pm$0.01 & 0.011$\pm$0.06 & 0.5$\pm$0.1 \\
%comptb  &                    &        &         &        &       &         &             \\
comptb   & $\alpha=\Gamma-1$ & 1.02$\pm$0.01  & 1.00$\pm$0.05    & 0.82$\pm$0.07  \\
         & kT$_{s}$ (keV)    & 1.36$\pm$0.04  & 1.39$\pm$0.07    & 2.0$\pm$0.1 \\
         & log A$$            & 1.29$\pm$0.03  &   0.86$\pm$0.03  &  0.74$\pm$0.03\\
         & kT$_{e}$ (keV)    & 13.98$\pm$0.08 &   9.62$\pm$0.06  &  10.3$\pm$0.2\\
         & N$_{Com}^{\dagger\dagger}$ & 7.47$\pm$0.05& 15.19$\pm$0.09 & 4.38$\pm$0.08\\
%Gaussian &               &       &         &              &         &         &           \\
Gaussian & E$_{line}$ (keV)  & 6.70$\pm$0.08 & 6.79$\pm$0.09  & 6.56$\pm$0.04\\
         & $\sigma_{line}$ (keV)& 0.65$\pm$0.05 & 0.60$\pm$0.01 & 0.60$\pm$0.01\\
         & N$_{line}^{\dagger\dagger}$ & 0.43$\pm$0.06 & 0.26$\pm$0.07 & 0.03$\pm$0.01\\
      \hline
         & $\chi_{red}^2$ (d.o.f.) & 3.2 (340)& 1.59 (421) & 12.3 (86) \\
      \hline
phabs     & N$_H$ (cm$^{-2}$)     & 2.9$\pm$0.1    & 3.76$\pm$0.02    & 5.7$^{frozen}$\\
bbody    & kT$_{BB}$ (keV)       & 0.56$\pm$0.07 & 0.53$\pm$0.02   & 0.60$\pm$0.07\\
         & N$_{BB}^{\dagger\dagger}$ & 0.084$\pm$0.001 &  0.012$\pm$0.004 & 0.001$\pm$0.001\\
% NS surface:
comptb1  &$\alpha_1=\Gamma_1-1$  & 1.00$\pm$0.03  & 1.01$\pm$0.07  & 0.97$\pm$0.02\\
         & kT$_{s1}$ (keV)       & 1.39$\pm$0.05  & 1.32$\pm$0.03  & 2.0$\pm$0.1 \\
         & log A$_1$              & 2.0$^{\dagger\dagger\dagger}$   & 2.0$^{\dagger\dagger\dagger}$ & 2.0$^{\dagger\dagger\dagger}$\\
         & kT$^{(1)}_{e}$ (keV)  & 15.3$\pm$0.4   & 8.47$\pm$0.08  & 9.7$\pm$0.1\\
         & N$_{Com1}^{\dagger\dagger}$&2.28$\pm$0.05&2.34$\pm$0.05 & 3.65$\pm$0.09\\
% CC/disc:   &                   &                &                &               &              &              &              \\
comptb2&$\alpha_2=\Gamma_2-1$    & 0.41$\pm$0.02  & 0.34$\pm$0.06  & 0.69$\pm$0.03\\
         & kT$_{s2}$ (keV)       & 0.83$\pm$0.06  & 0.80$\pm$0.09  & 1.05$\pm$0.08\\
%         & logA$_2$              & 2.0$^{\dagger}$ & 2.0$^{\dagger}$ & 2.0$^{\dagger}$ \\%&-0.28$\pm$0.05 & 2.0$^{\dagger}$ & 0.19$\pm$0.06\\
         & kT$^{(2)}_{e}$ (keV)  & 96$\pm$8       & 100$\pm$9      & 43$\pm$2\\
%         & logA$_2$              & 2.0$^{\dagger}$ & 2.0$^{\dagger}$ & 2.0$^{\dagger}$ &-0.28$\pm$0.05 & 2.0$^{\dagger}$ & 0.19$\pm$0.06\\
         & log A$_2$              & 2.0$^{\dagger\dagger\dagger}$   & 2.0$^{\dagger\dagger\dagger}$   & 0.39$\pm$0.04 \\ % &-0.28$\pm$0.05 & 2.0$^{\dagger}$ & 0.19$\pm$0.06\\
         & N$_{Com2}^{\dagger\dagger}$&0.14$\pm$0.07 & 0.5$\pm$0.1 & 0.8$\pm$0.1\\
Gaussian & E$_{line}$ (keV)      & 6.48$\pm$0.07  & 6.42$\pm$0.04  & 6.51$\pm$0.03\\
%     & $\sigma_{line}$ (keV)& 2.60$\pm$0.04  & 2.56$\pm$0.07& 2.79$\pm$0.08 & 2.64$\pm$0.07& 0.90$\pm$0.07& 1.54$\pm$0.06 \\
         & N$_{line}^{\dagger\dagger}$& 0.16$\pm$0.8  & 0.25$\pm$0.09 & 0.05$\pm$0.02\\
%$cyclabs$$^{\dagger\dagger\dagger\dagger}$  & E$_{cyc}$ (keV)  & 36.7(9) &  39.6(2)   & 39.2(1)   \\
%$cyclabs$  & E$_{cyc}$ (keV)  & 36.7(9) &  39.6(2)   & 39.2(1)   \\
%         & $\sigma_{cyc}$ (keV)  & 11.95$\pm$0.04 & 8.9$\pm$0.7 & 19.4$\pm$ 0.6\\
%         & Depth$_{cyc}$      & 0.4$\pm$ 0.1 & 0.3$\pm$ 0.1 &  0.5$\pm$ 0.1\\
%%Flux$^{\dagger\dagger\dagger}$ &    &            &         &       &       &        &        \\
%%     & 3 - 60 keV        & 5.18     & 5.48       & 5.01    & 4.76  & 7.79  & 5.51    \\
%%     & 13 - 150 keV      & 2.84     & 2.99       & 2.69    & 2.30  & 4.65  & 3.02    \\
      \hline
%   & $\chi_{red}^2$ (d.o.f.) & 0.97 (332)     & 1.10 (413)   & 1.11 (78)  \\
   & $\chi_{red}^2$ (d.o.f.) & 0.96 (338)     & 1.09 (419)   & 1.12 (83)  \\
      \enddata
  \label{tab:fit_table_SAX_Suzaku}
%      \label{tab:list_RXTE}
%    \label{tab:BeppoSAX_fit_table}
%\tablefoot{ 
$^\dagger$ Parameter errors are given at the 90\% confidence level.
$^{\dagger\dagger}$ %The spectral model is  $wabs*(blackbody + COMPTB + COMPTB + Gaussian)$,
The normalization parameters of Blackbody and Comptb components are in units of 
$L_{37}^{soft}/d^2_{10}$ $erg/s/kpc^2$, where $L_{37}^{soft}$ is the soft photon  luminosity in units of 10$^{37}$ erg/s, 
$d_{10}$ is the distance to the source in units of 10 kpc 
and Gaussian component is in units of $10^{-2}\times total~~photons$ $cm^{-2}s^{-1}$ in line;
$^{\dagger\dagger\dagger}$ parameter $\log(A_1)$ is fixed at 2.0  %(see comments in the text); 
%$^{\dagger\dagger}$ when parameter $\log(A)\gg1$, it is fixed to a value 2.0 
for the two-Comptb models; % $wabs*(Blackbody1+Blackbody2+Comptb+Gaussian)$, 
%parameter $\log(A_2)$ is fixed at 2.0 for the model $wabs*(Blackbody+Comptb1+Comptb2+Gaussian)$ (see comments in the text); 
$\sigma_{line}$ of Gaussian component is fixed to a value 0.6 keV for the model 
phabs*(Blackbody+\-Comptb1+Comptb2+Gaussian) (see comments in the text); 
$N_H$ is units of $10^{22}$ cm$^{-2}$. %}% (Iaria et al., 2006). 
\end{deluxetable}

\newpage

% &&& Table 5   &&&&&&&&&&&&&& ApJ format &&&&&&&&&&&&&&&&&&&&&&&&&&&&&

%
% TABLE 5 RXTE  ANALYSIS  sets R1-R3
%
\begin{deluxetable}{llcccccccccccccc}
\rotate
\tablewidth{0in}
\tabletypesize{\scriptsize}
    \tablecaption{Best-fit parameters of spectral analysis of {\it RXTE} 
%$Beppo$SAX and $Suzaku$ 
observations of 4U~1700-37 (R1 - R2 sets) in 3 -- 200~keV energy range$^{\dagger}$. 
Parameter errors correspond to 90\% confidence level.}
%\label{tab:fit_table_RXTE_1}
%{tab:fit_table_RXTE_1} 
    \renewcommand{\arraystretch}{1.2}
\tablehead{
Observational & MJD, & Exp &$\alpha_1=$  & $kT^{(1)}_e,$ & $N_{Com1}^{\dagger\dagger\dagger}$ & $kT_{s1}$, & $N_{Bbody}^{\dagger\dagger\dagger}$ & $kT_{s2}$, & $\alpha_2=$  & $kT^{(2)}_e,$ & $\log(A_2)$ & N$_{Com2}^{\dagger\dagger\dagger}$ & E$_{line}^{\dagger\dagger\dagger\dagger}$,& $N_{line}^{\dagger\dagger\dagger}$ &  $\chi^2_{red}$ \\ %& F$_1$/F$_2$/F$_3^{\dagger\dagger\dagger\dagger}$ \\
ObsId             & day  & ks &$\Gamma_1-1$ & keV           &            &                  &                                                       &  keV       & $\Gamma_2-1$ & keV           &            &                        & keV       &                                                &      (d.o.f.)                                       
%
%Observational & MJD, & $\alpha_1=$  & $kT^{(1)}_e,$ & $N_{Com1}^{\dagger\dagger\dagger}$ & $kT_{s1}$, & $N_{Bbody}^{\dagger\dagger\dagger}$ & $kT_{s2}$, & $\alpha_2=$  & $kT^{(2)}_e,$ & $\log(A_2)$ & N$_{Com2}^{\dagger\dagger\dagger}$ & E$_{line}^{\dagger\dagger\dagger\dagger}$,& $N_{line}^{\dagger\dagger\dagger}$ &  $\chi^2_{red}$ \\ %& F$_1$/F$_2$/F$_3^{\dagger\dagger\dagger\dagger}$ \\
%ID             & day  & $\Gamma_1-1$ & keV           &            &                  &                                                       &  keV       & $\Gamma_2-1$ & keV           &            &                        & keV       &                                                &      (d.o.f.)                                       
}
\startdata
% \startdata%   id     MJD    alf     T_e       log     norm_COMPTB    kT_bb     N_bb    E_line      N_line   Xi_2(dof)  Flux3-10 Fl10-60
% #%--------------------------------------------------------------------------------------------------------------------------------------------------------------------------------------------------------------------------------------------------------------------------
% #%N             MJD    alpha err_ T_e1,er_   log(A1)err  norm err_  kT_s1 err N_bb err  kTs(2)er_  alpha(2)er_kT_e(2)er logA(2)er_ norm(2)er E_ga(1) nor_err_ xi2(dof)       flux
% #%                        alpha     T_e1       logA      comptb1      kT_s1    _bb       kTs2       alph(2)   kT_e(2)  logA(2)  com  norm(2)   fe_ga    Fe                3-10/10-50/50-200/3-80  cnt er exp
% #------1----------2---------3----------4--------5------------------------6-----------7---------8-----------9--------10--------11-------12----------13---------14------15-----16----17----18----19------20-----21------22-----23-----24------25-----26-----27-----28-----29-----30----31-------32----33---34-----35--36-----37---38---39---40------41-42---43-------------------------------------------------------------------------------
10148-01-01-000  & 50334.29 & 11.759 & 0.99(2) &  10.28(7) & 5.73(2) & 1.42(3) & 0.53(4)  & 0.73(6) & 0.34(9) & 70(5)  & 0.36(4) & 0.8(2)  & 6.50(1) & 0.06(2) & 1.17(81) \\%2.80 5.86 0.87  9.28 86.6  2336
10148-01-02-00   & 50339.01 & 7.567  & 0.97(2) &   9.6(1)  & 3.62(9) & 2.0(1)  & 0.001(1) & 1.04(7) & 0.68(3) & 42(2)  & 0.39(5) & 0.9(1)  & 6.51(3) & 0.05(2) & 1.12(81) \\%1.05 3.91 0.65  5.37 59.6  1043
10148-01-03-000  & 50335.41 & 10.847 & 1.00(1) &  10.5(1)  & 4.40(3) & 1.80(2) & 0.001(1) & 0.81(1) & 0.37(5) & 74(8)  & 0.37(2) & 0.9(2)  & 6.54(5) & 0.06(2) & 1.24(81) \\%1.79 4.67 0.82  7.01 68.7  1569
30094-01-01-00   & 51285.53 & 11.359 & 1.07(4) &  8.8(1)   & 2.03(3) & 1.8(1)  & 0.002(1) & 0.96(2) & 1.02(4) & 23(2)  & 0.32(4) & 0.7(1)  & 6.53(2) & 0.04(2) & 0.91(81) \\%0.87 2.14 0.36  3.20 367.0 450.6
30094-01-02-00   & 51285.26 & 3.599  & 1.00(2) &  20.01(2) & 0.09(3) & 1.25(2) & 0.003(1) & 1.10(3) & 1.01(7) & 24(3)  & 0.34(3) & 0.6(1)  & 6.50(1) & 0.06(2) & 0.85(81) \\%  0.02 0.06  0.04  0.11   63.5  7.67
30094-01-03-00   & 51285.33 & 3.583 & 0.99(7) &  20.45(3) & 0.01(3) & 1.23(1) & 0.002(1) & 1.09(2) & 1.01(5) & 25(2)  & 0.36(4) & 0.03(1) & 6.41(1) & 0.05(2) & 0.87(81) \\%  0.01 0.01  0.02  0.02   0.2   5.2
30094-01-04-00   & 51285.39 & 3.567 & 1.05(4) &  19.01(2) & 0.19(3) & 1.24(3) & 0.003(2) & 1.12(3) & 1.03(7) & 27(4)  & 2.00$^{\dagger\dagger}$   & 0.03(2) & 6.42(2) & 0.05(1) & 0.85(80) \\%0.07 0.18  0.06  0.27   2.9   38.2
30094-01-05-00   & 51285.51 & 3.615 & 0.97(2) &  16.3(1)  & 0.74(5) & 1.25(2) & 0.003(1) & 1.13(4) & 1.02(3) & 20(7)  & 0.37(2) & 0.01(1) & 6.41(1) & 0.05(2) & 1.05(81) \\%0.08 0.76  0.26  0.84   12.9  48.2 %# N_H increase!
30094-01-06-00   & 51288.33 & 3.711 & 0.98(3) &  21.9(2)  & 0.25(3) & 1.23(4) & 0.004(3) & 1.11(3) & 1.01(3) & 20(9)  & 0.36(4) & 0.01(1) & 6.46(2) & 0.05(1) & 1.07(81) \\%0.09 0.25  0.16  0.40   11.9  72.3
30094-01-07-00   & 51288.39 & 3.599 & 0.99(5) &  24.2(1)  & 0.17(3) & 1.25(3) & 0.004(2) & 1.07(5) & 0.94(1) & 21(6)  & 0.09(2) & 0.01(1) & 6.46(2) & 0.26(1) & 1.22(81) \\%0.07 0.19  0.16  0.25   3.07  59.8
30094-01-08-00   & 51291.21 & 3.088 & 1.00(9) &  6.48(9)  & 1.46(2) & 1.54(2) & 0.001(1) & 1.10(2) & 1.00(5) & 22(1)  & -0.15(1)& 0.63(4) & 6.41(3) & 0.23(4) & 1.02(81) \\%0.60 1.33  0.20  2.04   23.4 309.0
30094-01-09-00   & 51291.26 & 3.775 & 1.01(8) &  9.8(1)   & 11.67(9)& 1.88(9) & 0.10(2)  & 1.01(1) & 0.51(3) & 52(8)  & -0.32(4)& 1.06(9) & 6.52(8) & 0.74(9) & 1.14(81) \\%3.07 11.53 2.31  15.93  193. 1813.0
30094-01-10-00   & 51291.33 & 3.152 & 1.02(3) &  20.0(8)  & 0.39(2) & 1.25(7) & 0.004(1) & 0.9(1)  & 0.6(2)  & 49(7)  & 2.00$^{\dagger\dagger}$   & 0.02(1) & 6.46(2) & 0.26(1) & 1.04(80) \\%0.13 0.40  0.29  0.64   7.4  72.5
30094-01-11-00   & 51291.59 & 6.815 & 0.99(5) &  16.1(1)  & 0.12(5) & 1.25(2) & 0.001(1) & 1.10(1) & 1.0(10) & 20(1)  & 0.35(6) & 0.03(1) & 6.39(7) & 0.03(1) & 0.90(81) \\%0.05 0.01  0.05  0.11   2.4  33.37
30094-01-12-00   & 51291.87 & 9.215 & 1.00(2) &  8.00(1)  & 0.09(3) & 2.01(4) & 0.001(1) & 0.89(4) & 0.40(8) & 90(10) & 2.00$^{\dagger\dagger}$   & 0.08(8) & 6.47(2) & 0.10(1) & 1.10(80) \\%0.04 0.16  0.46  0.30   4.9  31.04
30094-01-13-00   & 51292.07 & 1.152 & 1.01(4) &  18.0(3)  & 0.05(3) & 1.26(3) & 0.002(1) & 1.10(1) & 1.07(3) & 19(2)  & 2.00$^{\dagger\dagger}$   & 0.07(1) & 6.46(3) & 0.09(2) & 1.11(80) \\%0.04 0.18  0.55  0.32   1.3  20.1        
30094-01-14-00   & 51294.79 & 10.623 & 1.00(8) &  9.05(2)  & 1.39(3) & 2.05(1) & 0.010(1) & 0.96(3) & 0.65(4) & 52(3)  & -0.15(4)& 0.10(6) & 6.39(2) & 0.27(2) & 0.95(81) \\%0.29  1.38  0.23  1.82   23.9 251.5
30094-01-15-00   & 51298.06 & 1.360 & 1.01(7) &  15.87(6) & 3.32(3) & 1.76(4) & 0.020(3) & 0.86(3) & 0.40(6) & 69(5)  & 2.00$^{\dagger\dagger}$   & 0.09(3) & 6.38(4) & 0.24(3) & 1.06(80) \\%  0.29  1.41  0.13  1.83   65.7 770.6
30094-01-16-00   & 51298.13 & 1.599 & 0.89(5) &  16.5(2)  & 2.93(5) & 1.64(2) & 0.003(2) & 0.84(2) & 0.35(3) & 86(12) & 2.00$^{\dagger\dagger}$   & 0.09(5) & 6.39(3) & 0.25(3) & 0.89(80) \\%  0.84  3.26  1.88  4.88   58.3 643.8                                 
30094-01-17-00   & 51298.61 & 1.712 & 1.03(7) &  15.17(6) & 0.17(4) & 1.24(6) & 0.002(4) & 1.11(3) & 1.02(3) & 23(2)  & 0.37(8) & 0.03(2) & 6.42(2) & 0.032(4)& 1.18(81) \\%0.08 0.24  0.20  0.38   2.6   28.2
30094-01-18-00   & 51298.67 & 1.536 & 1.06(3) &  15.78(5) & 0.12(3) & 1.26(3) & 0.003(1) & 1.00(5) & 1.00(4) & 24(2)  & 0.52(6) & 0.02(1) & 6.42(3) & 0.013(3)& 0.76(81) \\%0.04 0.08  0.07  0.14   20.6  12.6
30094-01-20-00   & 51298.79 & 2.496 & 1.01(5) &  17.32(7) & 0.19(7) & 1.25(2) & 0.003(1) & 1.01(2) & 0.97(3) & 23(2)  & 0.71(4) & 0.03(2) & 6.45(8) & 0.014(3)& 0.83(81) \\%0.03 0.07  0.05  0.12   0.9   11.3
30094-01-21-00   & 51298.86 & 2.432 & 0.99(6) &  17.54(6) & 0.13(3) & 1.24(4) & 0.004(1) & 1.00(3) & 1.03(3) & 22(3)  & 0.46(3) & 0.02(1) & 6.50(7) & 0.015(3)& 1.35(81) \\%0.04 0.09  0.07  0.15   11.4  32.1
30094-01-22-00   & 51301.60 & 1.599 & 1.02(8) &  11.58(2) & 2.22(1) & 1.87(3) & 0.52(1)  & 0.83(2) & 0.20(3) & 90(15) & 2.00$^{\dagger\dagger}$   & 0.08(3) & 6.42(3) & 0.25(2) & 1.36(80) \\%  0.55  2.36 0.42  3.32  38.1 445.6
30094-01-23-00   & 51301.65 & 1.999 & 0.97(7) &  13.18(1) & 3.15(2) & 1.38(6) & 0.02(1)  & 0.81(6) & 0.25(3) & 86(11) & 2.00$^{\dagger\dagger}$   & 0.09(4) & 6.45(2) & 0.24(3) & 0.74(80) \\%  0.54  3.28 1.59  4.49  56.1 503.4
30094-01-24-00   & 51301.76 & 2.016 & 0.98(8) &  13.04(2) & 5.87(1) & 1.49(5) & 0.023(2) & 0.85(2) & 0.27(3) & 80(11) & 2.00$^{\dagger\dagger}$   & 0.07(3) & 6.52(3) & 0.10(3) & 0.85(80) \\%  0.87  5.91 2.15  7.87  103.4 651.9
30094-01-26-00   & 51301.87 & 2.304 & 1.04(7) &   7.89(8) & 0.20(6) & 2.1(2)  & 0.004(1) & 1.00(3) & 1.07(3) & 29(2)  & -0.22(4)& 0.27(2) & 6.46(2) & 0.39(4) & 1.33(81) \\%0.10 0.29  0.12  0.43   4.5   52.8
30094-01-27-00   & 51309.39 & 1.728 & 1.02(9) &  12.05(2) & 0.90(5) & 1.23(4) & 0.002(1) & 1.10(5) & 1.00(3) & 23(14) & 0.35(2) & 0.10(3) & 6.52(3) & 0.05(2) & 0.99(81) \\%0.13 0.93  0.29  1.26  13.6 136.4
30094-01-28-00   & 51309.52 & 2.479 & 1.04(4) &  14.19(3) & 4.18(2) & 1.10(1) & 0.002(1) & 0.85(2) & 0.34(3) & 84(15) & 2.00$^{\dagger\dagger}$   & 0.009(1)& 6.40(4) & 0.25(6) & 1.12(80) \\%  0.97  4.47  1.96  6.34   77.4  609.1
30094-01-29-00   & 51309.58 & 2.672 & 0.94(6) &   9.73(2) & 2.32(3) & 1.68(2) & 0.001(1) & 0.81(2) & 0.29(3) & 91(10) & 2.00$^{\dagger\dagger}$   & 0.30(8) & 6.39(2) & 0.24(4) & 0.94(80) \\%  0.86  2.42  1.23  3.63   43.4  459.9
30094-01-30-00   & 51326.9  & 1.792 & 0.97(9) &   8.37(2) & 0.15(7) & 2.0(2)  & 0.50(1)  & 1.09(3) & 0.79(3) & 45(4)  & 0.7(1)  & 0.11(7) & 6.43(3) & 0.38(9) & 1.08(81) \\%0.05 0.26  0.21  0.38  -9.4   33.6
30094-01-31-00   & 54124.9  & 1.792 & 0.99(3) &   8.50(4) & 1.41(9) & 2.12(6) & 0.020(3) & 0.79(5) & 0.20(3) & 100(20)& 2.00$^{\dagger\dagger}$   & 0.08(3) & 6.41(2) & 0.24(4) & 1.19(74) \\%  0.34 1.49  1.01  2.06  10.6  289.6
30094-01-32-00   & 54124.9  & 1.728 & 0.98(7) &   8.37(3) & 0.19(3) & 2.0(1)  & 1.001(1) & 1.09(3) & 1.01(3) & 23(4)  & 0.70(4) & 0.11(7) & 6.43(4) & 0.38(2) & 1.23(81) \\%0.13 0.95  0.51  1.29  13.6  136.4
30094-01-33-00   & 51407.70 & 3.392 & 1.02(2) &   8.37(2) & 0.21(7) & 1.28(7) & 0.001(1) & 1.09(2) & 0.70(3) & 67(12) & -0.4(1) & 0.14(3) & 6.41(3) & 0.17(1) & 1.13(81) \\%0.07 0.19  0.01(3) & 18.7  38.3
30094-01-34-00   & 51407.37 & 1.712 & 0.97(6) &   8.36(5) & 0.01(3) & 2.02(4) & 0.005(2) & 1.02(1) & 0.62(2) & 79(14) & 2.00$^{\dagger\dagger}$   & 0.03(3) & 6.42(5) & 0.04(1) & 0.89(80) \\%0.02 0.06  0.09   0.11  0.03  15.8
      \enddata
  \label{tab:fit_table_RXTE_1}
%\tablefoot{ 
$^\dagger$ The spectral model is  %phabs*cyc*(blackbody + \-Comptb1 + Comptb2 + Gaussian) for 1996 data, and 
phabs*(blackbody + \-Comptb1 + Comptb2 + Gaussian); % for the rest data; % where $N_H$ is fixed at 
%a value 9.6$\times 10^{22}$ cm$^{-2}$ (Church et al., 2006); % (Oosterbroek et al., 2001)
%a value 6.8$\times 10^{22}$ cm$^{-2}$ (Oosterbroek et al., 2001);  
color temperature %$T_s$ and 
$T_{BB}$ of Bbody component is fixed at %1.3 and 
0.6 keV % and ``seed'' photon temperatures $T_{s1}$/$T_{s2}$ are fixed at 1.1/1.5 keV, respectively
 (see comments in the text); 
$^{\dagger\dagger}$ parameter $\log(A_1)$ is fixed at 2.0 and %(see comments in the text), 
%$^{\dagger\dagger}$ 
when parameter $\log(A_2)\gg1$, this parameter is fixed at 2.0 (see comments in the text), 
$^{\dagger\dagger\dagger}$ normalization parameters of $blackbody$ and $COMPTB$ components are in units of 
$L_{37}/d^2_{10}$, where $L_{37}$ is the source luminosity in units of 10$^{37}$ erg/s, 
$d_{10}$ is the distance to the source in units of 10 kpc 
and Gaussian component is in units of $10^{-2}\times total~~photons$ $cm^{-2}s^{-1}$ in line;   
%$^{\dagger\dagger\dagger\dagger}$ $kT_{s1}$ varies  from 1.3 keV to 1.5 keV for  
%the {\it island}  to banana state transition while it is concentrated  about 1.5 keV (see Fig.~\ref{index_temperature_s_12}) for most of the spectra spectra,
$^{\dagger\dagger\dagger\dagger}$ $\sigma_{line}$ of Gaussian component is fixed to a value 0.7 keV (see comments in the text),
%$N_H$ was fixed at value of 2.4$\times 10^{22}$ cm$^{-2}$ (BO02). %, %Iaria et al., 2006),  
$N_H$ was free to vary within the range of (2 -- 8)$\times 10^{22}$ cm$^{-2}$ (see comments in the text). %; 
%energy, width and depth of cyclotron absorption features for 1996 data are varied in 36 -- 39 keV, 11  -- 19 keV and 0.1 -- 0.4 ranges, respectively. % (BO02). %, %Iaria et al., 2006),  
%$^{\dagger\dagger\dagger\dagger}$spectral fluxes (F$_1$/F$_2$/F$_2$) in units of $\times 10^{-9}$ ergs/s/cm$^2$ for  (3 -- 10), (10 -- 50) keV and (50 -- 200) keV energy ranges respectively.  
%in units of $\times 10^{-9}$ ergs/s/cm$^2$.
%* this observations are  fitted with $bmc+Gaussian1+Gaussian2+bbody$ model, see values of the best-fit BB color temperature and EW in Table 2, 3 and 4.
%$^{\dagger\dagger}$ parameter $\log(A_2)$ is fixed at 2.0 (see comments in the text), 
%}
\end{deluxetable}

\newpage

%&&&&& Table 6  &&&&&&&&&&&&&& ApJ format &&&&&&&&&&&&&&&&&&&&&&&&&&&&&

%
% TABLE 6 RXTE  ANALYSIS  sets R4-R5
%

\begin{deluxetable}{llcccccccccccccc}
\rotate
\tablewidth{0in}
\tabletypesize{\scriptsize}
    \tablecaption{Best-fit parameters of spectral analysis of 
%$Beppo$SAX and $Suzaku$  
{\it RXTE} observations (R3-R5 sets) of 4U~1700-37 in 3 -- 200~keV 
energy range$^{\dagger}$. Parameter errors correspond to 90\% confidence level.}
%\label{tab:fit_table_RXTE_2}
    \renewcommand{\arraystretch}{1.2}
\tablehead{
Observational & MJD, & Exp &$\alpha_1=$  & $kT^{(1)}_e,$ & $N_{Com1}^{\dagger\dagger\dagger}$ & $kT_{s1}$, & $N_{Bbody}^{\dagger\dagger\dagger}$ & $kT_{s2}$, & $\alpha_2=$  & $kT^{(2)}_e,$ & $\log(A_2)$ & N$_{Com2}^{\dagger\dagger\dagger}$ & E$_{line}^{\dagger\dagger\dagger\dagger}$,& $N_{line}^{\dagger\dagger\dagger}$ &  $\chi^2_{red}$ \\ %& F$_1$/F$_2$/F$_3^{\dagger\dagger\dagger\dagger}$ \\
ObsId             & day  & ks &$\Gamma_1-1$ & keV           &            &                  &                                                       &  keV       & $\Gamma_2-1$ & keV           &            &                        & keV       &                                                &      (d.o.f.)                                       
%
%Observational & MJD, & $\alpha_1=$  & $kT^{(1)}_e,$ & $N_{Com1}^{\dagger\dagger\dagger}$ & $kT_{s1}$, & $N_{Bbody}^{\dagger\dagger\dagger}$ & $kT_{s2}$, & $\alpha_2=$  & $kT^{(2)}_e,$ & $\log(A_2)$ & N$_{Com2}^{\dagger\dagger\dagger}$ & E$_{line}^{\dagger\dagger\dagger\dagger}$,& $N_{line}^{\dagger\dagger\dagger}$ &  $\chi^2_{red}$ \\ %& F$_1$/F$_2$/F$_3^{\dagger\dagger\dagger\dagger}$ \\
%ID             & day  & $\Gamma_1-1$ & keV           &            &                  &                                                       &  keV       & $\Gamma_2-1$ & keV           &            &                        & keV       &                                                &      (d.o.f.)                                       
}
\startdata
30095-01-01-00   & 51129.66 & 3.647 & 1.04(3) &  12.89(2) & 21.7(1) & 2.03(1) & 0.11(8)  & 0.89(5) & 0.5(1)  & 72(19) & -0.39(5)& 1.03(6) & 6.48(2) & 2.0(2)  & 1.19(81) \\%8.63 23.42 5.54  35.76  416  7545
30095-01-01-01   & 51130.27 & 3.023 & 1.01(8) &  15.81(8) & 2.62(2) & 1.48(2) & 0.020(3) & 0.82(2) & 0.35(9) & 80(20) & -0.07(3)& 0.10(3) & 6.42(1) & 0.24(5) & 1.19(81) \\%1.38 3.00  1.18  5.02  52.2  682.2
30095-01-01-02   & 51129.23 & 3.455 & 0.99(9) &  12.87(1) & 0.02(4) & 2.08(1) & 0.002(3) & 0.96(4) & 0.59(8) & 60(15) & -0.15(2)& 0.20(4) & 6.46(2) & 0.85(4) & 1.18(81) \\%2.11 5.90  1.38  8.95  104.6  1852
30095-01-01-03   & 51129.73 & 3.695 & 1.05(4) &  10.12(9) & 7.20(5) & 1.95(4) & 0.12(1)  & 0.89(3) & 0.5(2)  & 73(18) & -0.48(7)& 1.02(8) & 6.40(3) & 0.81(7) & 0.82(81) \\%3.59 7.61  1.72  12.11 129.4 2972
30095-01-01-04   & 51129.80 & 3.695 & 1.09(7) &   9.94(3) & 6.63(8) & 1.37(2) & 0.93(4)  & 0.91(2) & 0.59(7) & 67(21) & -0.11(8)& 0.20(9) & 6.41(5) & 0.80(9) & 0.89(81) \\%3.26 6.54  0.90  10.42 107.0 2673
30095-01-01-05   & 51129.86 & 5.120 & 1.10(6) &  11.58(2) & 6.63(7) & 1.93(5) & 0.002(6) & 0.95(5) & 0.69(8) & 69(20) & -0.12(7)& 0.21(7) & 6.41(3) & 0.96(7) & 0.96(81) \\%3.07 7.03  1.29  11.00 123.1 2578
30095-02-01-04   & 51114.63 & 1.135 & 0.89(4) &  10.11(9) & 6.54(3) & 1.49(4) & 0.51(3)  & 1.09(2) & 0.93(8) & 27(3)  & -0.15(3)& 0.32(4) & 6.42(2) & 0.73(8) & 1.19(81) \\%3.27 6.57  0.82  10.48  107.0  2673
30095-02-01-05   & 51114.56 & 3.455 & 0.98(7) &  10.16(8) & 6.56(3) & 1.47(3) & 0.50(2)  & 1.06(2) & 0.95(7) & 30(6)  & -0.14(3)& 0.30(2) & 6.42(5) & 0.72(9) & 1.18(81) \\%3.27 6.57  0.82  10.48  107.0  2673
30095-02-02-00   & 51193.01 & 12.671 & 1.01(8) &   8.37(2) & 1.03(3) & 2.0(1)  & 0.13(1)  & 1.09(3) & 0.89(3) & 30(4)  & 2.00$^{\dagger\dagger}$   & 0.05(7) & 6.49(2) & 0.09(1) & 1.17(80) \\%0.04  0.08  0.04  0.13  0.97 55.4%30095-02-02-01   & 51193.21 & 3.503 & 0.95(6) &   8.35(6) & 0.11(4) & 2.0(2)  & 0.11(6)  & 1.08(2) & 1.01(2) & 26(3)  & 2.00$^{\dagger\dagger}$   & 0.06(8) & 6.50(3) & 0.02(1) & 1.14(80) \\%0.06  0.16  0.05  0.24  1.8  34.01
30095-02-02-01   & 51193.21 & 3.503 & 0.95(6) &   8.35(6) & 0.11(4) & 2.0(2)  & 0.11(6)  & 1.08(2) & 1.01(2) & 26(3)  & 2.00$^{\dagger\dagger}$   & 0.06(8) & 6.50(3) & 0.02(1) & 1.14(80) \\%0.06  0.16  0.05  0.24  1.8  34.01
30095-02-02-02   & 51193.39 & 3.312 & 0.98(2) &  10.22(3) & 2.94(2) & 1.49(2) & 0.028(1) & 0.78(8) & 0.20(3) & 100(20)& -0.56(9)& 0.30(4) & 6.42(2) & 0.41(5) & 1.12(81) \\%0.51  2.83  0.98  3.71  47.26  583.2
30095-02-02-03   & 51193.39 & 6.527 & 1.01(8) &  22.59(2) & 0.43(3) & 2.04(3) & 0.027(1) & 1.08(3) & 0.98(4) & 23(3)  & 0.35(4) & 0.14(5) & 6.42(8) & 0.10(3) & 1.11(81) \\%0.07  0.50  0.27  0.71  9.98   97.69

40068-01-01-01   & 51777.13 & 1.888 & 1.03(5) &  10.84(9) & 5.60(2) & 1.53(6) & 0.92(3)  & 1.53(6) & 1.00(3) & 26(9)  & 2.00$^{\dagger\dagger}$   & 0.70(4) & 6.97(3) & 0.24(1) & 0.83(80) \\%2.34  6.25  1.37  9.41  10.75  820.6
40068-01-01-02   & 51778.19 & 13.871 & 1.02(7) &   9.84(7) & 6.75(6) & 1.30(2) & 0.01(1)  & 0.81(3) & 0.31(2) & 69(12) & 2.00$^{\dagger\dagger}$   & 0.10(3) & 6.51(7) & 0.54(9) & 1.15(80) \\%3.22  6.69  1.54  10.67 107.5  820.6
40068-01-01-03   & 51777.20 & 1.648 & 1.06(2) &   6.93(1) & 1.9(1)  & 1.15(9) & 0.001(3) & 1.26(4) & 1.00(3) & 21(6)  & 2.00$^{\dagger\dagger}$   & 0.70(4) & 6.40(3) & 0.21(6) & 1.14(80) \\%1.45 2.33  0.54  4.05  115.5  536.0
40068-01-01-04   & 51779.05 & 1.456 & 1.02(8) &  10.72(3) & 4.62(4) & 1.8(1)  & 0.05(1)  & 0.8(1)  & 0.30(3) & 90(25) & -0.54(7)& 0.40(1) & 6.47(8) & 0.53(7) & 1.16(81) \\%1.22 4.64  1.26  6.47  40.8  457.4
40068-01-01-05   & 51778.12 & 1.856 & 0.99(7) &  10.53(1) & 6.92(2) & 1.67(3) & 0.02(3)  & 0.83(2) & 0.34(7) & 83(20) & -0.40(3)& 0.10(3) & 6.39(3) & 0.10(2) & 1.19(81) \\%3.18 7.02  0.93  1.09  79.2  479.5
40068-01-01-06   & 51779.12 & 1.888 & 1.01(2) &   9.05(2) & 3.69(4) & 2.07(5) & 0.03(1)  & 1.09(7) & 0.9(1)  & 25(4)  & -0.15(3)& 2.07(3) & 6.38(2) & 0.15(8) & 1.17(81) \\%1.07 4.22  1.24  5.86  118.6  1053
40068-01-01-07   & 51779.18 & 2.272 & 1.07(7) &  12.30(6) & 1.28(3) & 1.65(4) & 0.02(2)  & 0.81(2) & 0.36(9) & 74(23) & -0.40(3)& 0.10(4) & 6.37(8) & 0.10(2) & 1.19(81) \\%0.36 1.33  0.35  1.90  71.9  429.1
40068-01-01-08   & 51778.05 & 1.440 & 1.02(3) &  10.74(4) & 2.20(9) & 1.71(5) & 0.001(3) & 0.79(3) & 0.3(2)  & 81(20) & 2.00$^{\dagger\dagger}$   & 0.70(6) & 6.40(7) & 0.33(5) & 1.13(80) \\%1.39 2.99  0.44  4.74  50.8  455.8
40068-01-01-09   & 51777.85 & 1.616 & 1.01(6) &   8.53(7) & 4.12(8) & 1.56(4) & 0.001(3) & 0.82(4) & 0.34(5) & 73(20) & 0.46(5) & 0.63(3) & 6.41(6) & 0.27(8) & 1.12(81) \\%2.54 4.41  0.73  7.34  74.8  409.6
40068-01-01-10   & 51777.79 & 2.896 & 1.00(7) &   1.71(3) & 22.68(3)& 2.05(7) & 0.002(3) & 1.09(2) & 1.01(3) & 23(2)  & 0.35(4) & 2.0(1)  & 6.40(8) & 0.05(2) & 0.84(81) \\%0.87 4.22  1.06  5.65   70.80 383.5
40068-01-01-011G & 51777.26 & 11.615 & 1.02(8) &   9.94(7) & 9.94(2) & 2.12(6) & 0.002(3) & 1.08(3) & 1.04(3) & 19(1)  & 0.31(7) & 3.0(2)  & 6.43(2) & 0.10(8) & 1.13(81) \\%4.05 9.40  2.11  14.72  166.7 691.2
40068-01-01-14   & 51778.98 & 1.008 & 1.03(5) &   7.02(3) & 1.2(1)  & 1.75(7) & 0.001(3) & 1.10(2) & 1.0(2)  & 19(3)  & 2.00$^{\dagger\dagger}$   & 0.55(6) & 6.45(9) & 0.18(5) & 1.11(80) \\%0.57 1.53  0.45  2.25  26.91  204.1
40068-01-01-15   & 51779.25 & 2.672 & 1.05(3) &  10.55(9) & 3.23(4) & 1.84(3) & 0.001(3) & 0.80(3) & 0.20(3) & 100(20)& -0.14(3)& 0.12(3) & 6.41(8) & 0.31(4) & 1.16(81) \\%0.89 3.22  0.81  4.50  56.54  174.2
40068-01-01-16   & 51777.52 & 2.768 & 1.02(5) &  10.83(1) & 6.88(3) & 1.42(6) & 0.51(1)  & 0.80(4) & 0.29(3) & 90(9)  & -0.35(4)& 0.09(4) & 6.40(3) & 0.69(8) & 0.93(81) \\%3.37 6.95  1.15  11.09  56.54  174.2
40068-01-01-17   & 51777.59 & 9.999 & 0.98(8) &  10.12(3) & 3.85(2) & 1.54(2) & 0.52(2)  & 0.91(2) & 0.51(7) & 52(4)  & 2.00$^{\dagger\dagger}$   & 0.08(2) & 6.42(7) & 0.38(6) & 1.08(80) \\%1.71 3.91  0.77  6.05   64.47  291.1

80105-14-01-00   & 52890.09 & 9.535 & 1.00(2) &   9.47(5) & 2.35(4) & 1.39(3) & 0.50(1)  & 1.04(3) & 0.91(2) & 27(2)  & 2.00$^{\dagger\dagger}$   & 0.3(1)  & 6.49(2) & 0.33(5) & 1.15(80) \\%  1.17 2.56  0.56   4.01  43.5  194.6
      \enddata
 \label{tab:fit_table_RXTE_2}
%\tablefoot{ 
%$^\dagger$ The spectral model is  phabs*cyc*(blackbody + \-Comptb1 + Comptb2 + Gaussian); % where $N_H$ is fixed at 
$^\dagger$ The spectral model is  phabs*(blackbody + \-Comptb1 + Comptb2 + Gaussian); % where $N_H$ is fixed at 
%a value 9.6$\times 10^{22}$ cm$^{-2}$ (Church et al., 2006); % (Oosterbroek et al., 2001)
%a value 6.8$\times 10^{22}$ cm$^{-2}$ (Oosterbroek et al., 2001);  
color temperature %$T_s$ and 
$T_{BB}$ of Bbody component is fixed at %1.3 and 
0.6 keV % and ``seed'' photon temperatures $T_{s1}$/$T_{s2}$ are fixed at 1.1/1.5 keV, respectively
 (see comments in the text); 
$^{\dagger\dagger}$ parameter $\log(A_1)$ is fixed at 2.0 and %(see comments in the text), 
%$^{\dagger\dagger}$ 
when parameter $\log(A_2)\gg1$, this parameter is fixed at 2.0 (see comments in the text), 
$^{\dagger\dagger\dagger}$ normalization parameters of $blackbody$ and $COMPTB$ components are in units of 
$L_{37}/d^2_{10}$, where $L_{37}$ is the source luminosity in units of 10$^{37}$ erg/s, 
$d_{10}$ is the distance to the source in units of 10 kpc 
and Gaussian component is in units of $10^{-2}\times total~~photons$ $cm^{-2}s^{-1}$ in line;   
%$^{\dagger\dagger\dagger\dagger}$ $kT_{s1}$ varies  from 1.3 keV to 1.5 keV for  
%the {\it island}  to banana state transition while it is concentrated  about 1.5 keV (see Fig.~\ref{index_temperature_s_12}) for most of the spectra spectra,
$^{\dagger\dagger\dagger\dagger}$ $\sigma_{line}$ of Gaussian component is fixed to a value of 0.7 keV (see comments in the text),
%$N_H$ was fixed at value of 2.4$\times 10^{22}$ cm$^{-2}$ (BO02). %, %Iaria et al., 2006),  
$N_H$ was free to vary within the range of (2 -- 8)$\times 10^{22}$ cm$^{-2}$ (see comments in the text). %; 
%energy, width and depth of cyclotron absorption features are varied in 36 --39 keV, 11  -- 19 keV and 0.1 -- 0.4 ranges, 
%respectively. % (BO02). %, %Iaria et al., 2006),  
%$^{\dagger\dagger\dagger\dagger}$spectral fluxes (F$_1$/F$_2$/F$_2$) in units of $\times 10^{-9}$ ergs/s/cm$^2$ for  (3 -- 10), (10 -- 50) keV and (50 -- 200) keV energy ranges respectively.  
%in units of $\times 10^{-9}$ ergs/s/cm$^2$.
%* this observations are  fitted with $bmc+Gaussian1+Gaussian2+bbody$ model, see values of the best-fit BB color temperature and EW in Table 2, 3 and 4.
%$^{\dagger\dagger}$ parameter $\log(A_2)$ is fixed at 2.0 (see comments in the text), 
%}
\end{deluxetable}

\newpage

% Table 7  &&&&&&&&&&&&&& ApJ format &&&&&&&&&&&&&&&&&&&&&&&&&&&&&

\begin{deluxetable}{llccccccc}
%\rotate
\tablewidth{0in}
\tabletypesize{\scriptsize}
    \tablecaption{Comparisons of the best-fit parameters  of HMXB source 4U~1700-37 and LMXBs: {\it Z}-sources Sco~X-1$^{(1)}$  
and GX~340+0$^{(2)}$ and {\it atoll} sources GX~3+1$^{(3)}$, 4U~1728-34$^{(4)}$, 4U~1820-30$^{(5)}$, 4U~1705-44$^{(6)}$ and  
``$atoll$+{\it Z}'' source XTE~J1701-462$^{(7)}$} 
    \label{tab:par_comparison}
    \renewcommand{\arraystretch}{1.2}
 \tablehead
{Source & Alternative & Class$^{(8)}$& Distance, & Presence of & $kT_e$,  & $ N_{Com}$ &  $kT_{s}$  & $f$ \\
  name  & name        &          & kpc       & kHz QPO     & keV         &  $L_{39}^{soft}/{D^2_{10}}$        & keV  &  }
 \startdata%   id     MJD      kT_Bbody     N_bb     [kT_s]    alf       T_e       log    norm_COMPTB E_line[Sigma_l] N_line Xi_2(dof)  Flux3-10 Fl10-60
Sco~X-1     & V818 Sco & Z, Sp, B     & 2.8$^{(9)}$ &    +$^{(10)}$        & 3-180 &  0.3-3.4 & 0.4-1.8 & 0.08-1   \\
4U~1642-45  & GX 340+0 & Z, Sp, B     & 10.5$^{(11)}$ &    +$^{(14)}$        & 3-21   &  0.08-0.2  & 1.1-1.5 & 0.01-0.5   \\
4U~1744-26  & GX 3+1   & Atoll, Sp, B & 4.5$^{(12)}$ &     none$^{(15)}$      & 2.3-4.5 &  0.04-0.15 & 1.16-1.7 & 0.2-0.9   \\
4U~1728-34  & GX~354-0 & Atoll, Su, D & 4.2-6.4$^{(13)}$ & +$^{(16)}$&  2.5-15& 0.02-0.09 &  1.3 & 0.5-1\\     
4U~1820-30  &   ...    & Atoll, Su, B & 5.8-8$^{(17)}$& +$^{(18)}$&  2.9-21& 0.02-0.14 &  1.1-1.7 & 0.2-1\\     
XTE~J1701-462&  ...    & Atoll+Z, Su, D & 8.8$^{(19)}$& +$^{(20)}$&  ...   & ...       &  1-2.7 & ...\\     
4U~1705-44  &   ...    & Atoll, Sp, B & 7.4$^{(21)}$  &    +$^{(22)}$ & 2.7-100 &  0.01-0.08 & 1.1-1.5 & 0.2-1   \\
4U~1700-37  & HD153919 & HMXB, Sp, B  & 1.9$^{(23)}$  &    none$^{(24)}$ & 2-100 &  0.001-0.22 & 1.3-2.1 & 0.1-1   \\
      \enddata%     \hline
%\tablefoot{ 
\\References:
(1) TSS14; 
(2) STF13;
(3) ST12; 
(4) ST11;  
(5) TSF13; 
(6) STSS15; 
(7) LRH09;
(8) Classification of the system in the various schemes (see text): Sp = supercritical, Su = subcritical, 
B = bulge, D = disk; 
(9) Bradshaw et al. (1999); %\cite{Bradshaw99}; 
(10) Zhang et al. (2006); 
(11) Fender \& Henry  (2000), Ford et al. (1998),  Christian \& Swank (1997); 
(12) \citet{kk00}, Ford et al. (2000);  
(13) \citet{par78};  
(14) \citet{Jonker98};  
(15) \citet{stroh98}; 
(16) \citet{to99};  
(17) \citet{ST04};  
(18) \citet{Smale97}; 
(19) \citet{Lin07a}, \citet{lin09}; 
(20) \citet{Sana10}; 
(21) \citet{Haberl_Titarchuk95};  %Haberl \& Titarchuk, 1995, 
(22) BO02;
(23) Ankay et al. (2001); 
(24) Borozon et al. (2003)
%}
\end{deluxetable}

\newpage

%
% 
% Figure 1
%

% 
% Figure 2
%

\newpage

% 
%  FIgure 2 SAX Suzaku spectra
%

\newpage

% 
%  FIgure 3 - MODEL CHOICE RXTE
%

\newpage

% 
%  FIgure 5 six RXTE SPECTRA
%

  \begin{figure}
% \centering
%    \includegraphics[width=8.5cm]{f5_11_1700_up.eps}
    \includegraphics[width=8.5cm]{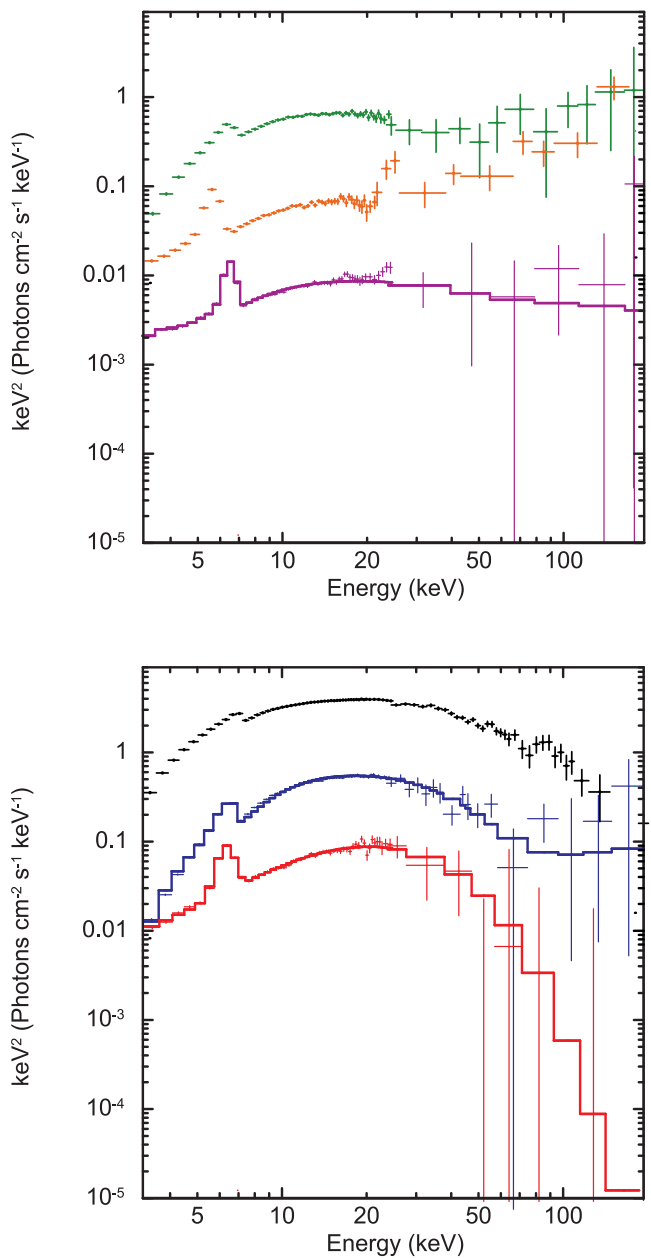}
      \caption{
Six representative $EF_E$ spectral diagrams for the hard (upper panel) and soft  
(lower panel) spectral states of 4U~1700-37. Data are taken from {\it RXTE} observations
%Typical $hard$ state spectra ($left$ panel): 
30094-01-01-10 (green), %($\varphi=0.7$), 
30094-01-12-00 (orange), %($\varphi=0.5$), 
30095-02-02-20 (violet), %($\varphi=0.6$), 
%and typical $soft$ state spectra ($right$ panel): 
30094-01-33-00 (red), %($\varphi=0.4$), 
30094-01-31-00 (blue), % ($\varphi=0.7$)
30095-01-01-00 (black). %($\varphi=0.6$).
}
   \label{6_spectra_rxte}
 \end{figure}

\newpage

% 
%  FIgure 6 - PHOTON INDEX
%

  \begin{figure}
% \centering
%    \includegraphics[width=17cm]{f6_1700.eps}
    \includegraphics[width=17cm]{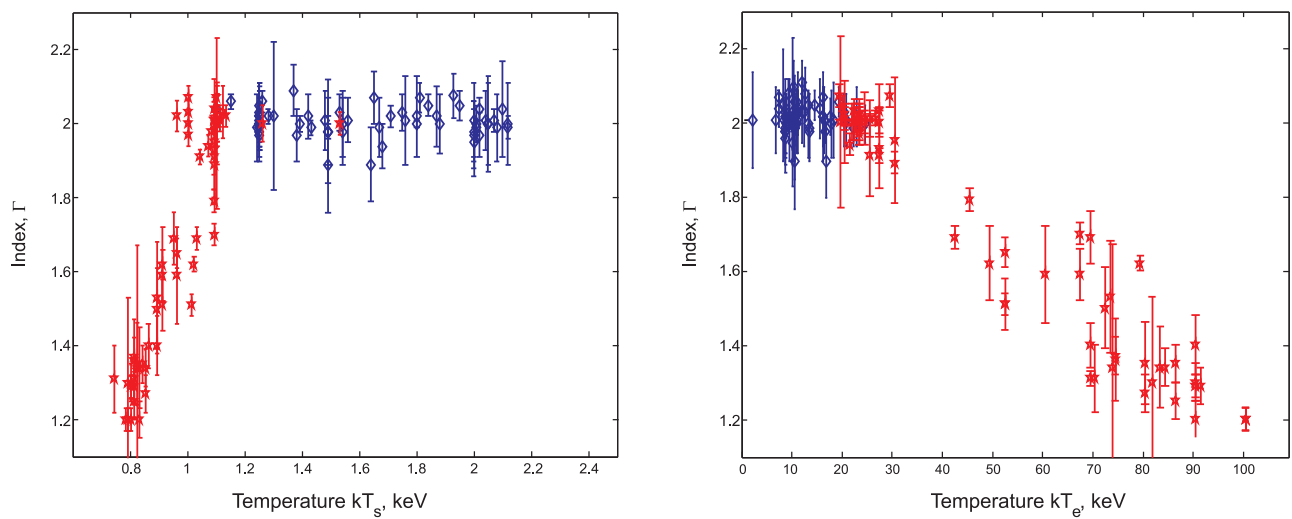}
      \caption{
The photon indices $\Gamma_1$ and $\Gamma_2$ plotted %vs 
versus the best-fit seed photon $kT_s$ (left) and electron $kT_e$ (right) 
%seed photon 
temperatures of the Comptb1 and Comptb2 components, respectively,  measured in keV  
% in the frame of  our spectral model $wabs*(blackbody+Comptb1+Comptb2+Gaussian)$ during RXTE %observations 
(see also  Tables 5-6). 
Blue and red points %($bottom$ and $top$ panels) 
correspond to Comptb1 and Comptb2 components, respectively, which are related to 
the thermal Comptonization of the soft  photons by plasma (electrons) in the Compton cloud.
% and CC, respectively. 
%The range of the high electron temperatures $kT^{(2)}_{e}$ where $\Gamma_2<2$ is shaded by $hazel$ color.}
}
\label{two_state_spectra}
\end{figure}

\newpage

% 
%  FIgure 7 - EVOLUTION lc
%

\newpage
% 
% Figure 8
%
\begin{figure}[ptbptbptb]
\includegraphics[scale=1.1,angle=0]{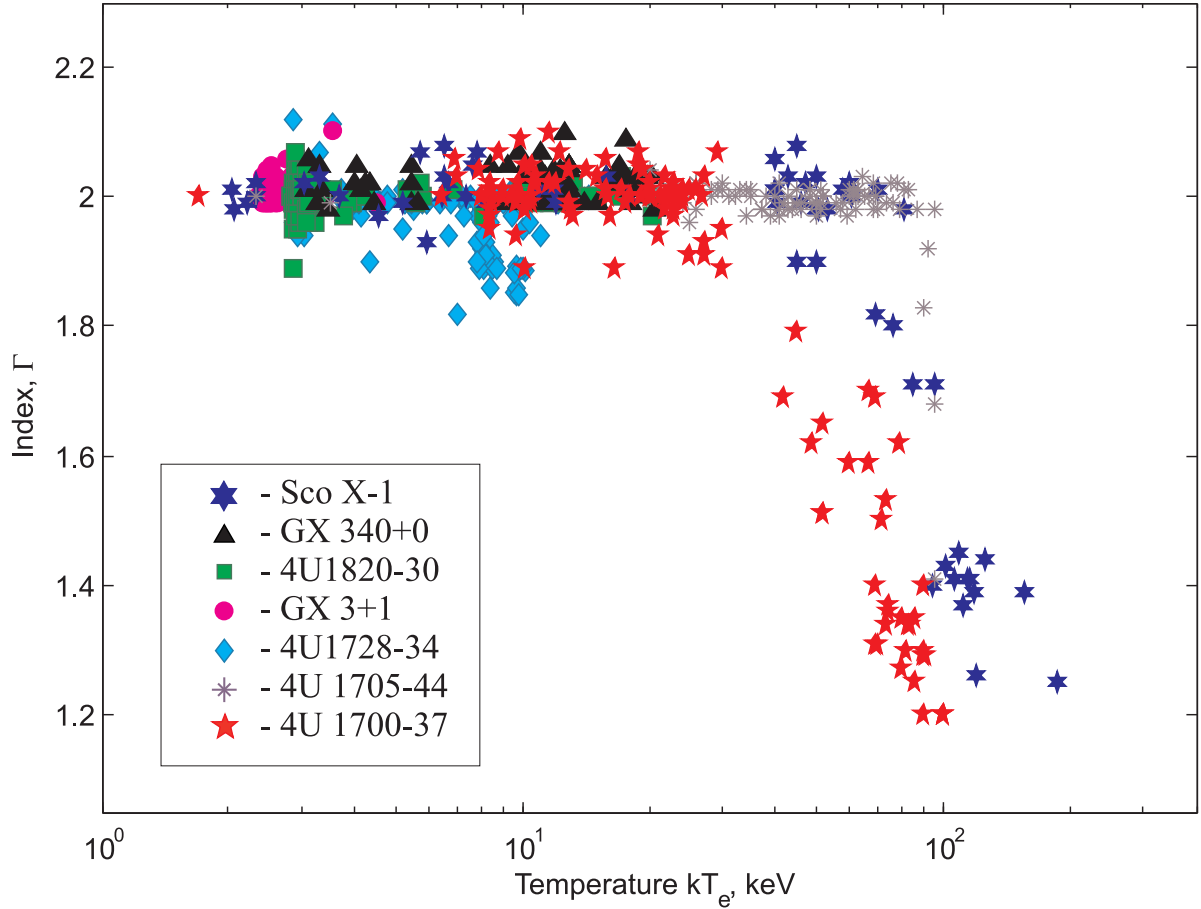}
\caption{The photon index $\Gamma$ vs $kT_e$ for  Z-sources Sco~X-1 (blue
stars, TSS14)), 
GX~340+0 (black triangles, STF13) and atoll sources 4U~1705-44 (grey, STSS15),
4U~1728-34 
(bright~blue diamonds, ST11), GX~3+1 (pink circles, ST12) and 
4U~1820-30 (green squares, TSF13), and NS 4U~1700-37 (red stars). 
}
\label{gam_te_7obj}
\end{figure}
\end{document}